\documentclass[aps, prl, twocolumn, superscriptaddress, floatfix]{revtex4-2}
\usepackage{graphicx}
\usepackage{amsmath}
\usepackage{ulem}
\usepackage{amssymb}
\usepackage{tikz}
\usepackage{braket}
\usepackage{float}
\usepackage{pgfplots}
\usepackage{mathrsfs}
\usepackage[version=4]{mhchem}
\pgfplotsset{compat=1.17}
\usepackage{subcaption}
\usetikzlibrary{shapes.geometric}
\usetikzlibrary{shadows,patterns,perspective}
\usetikzlibrary {decorations,decorations.text}
\usetikzlibrary{decorations.pathreplacing}
\usepackage{float}
\usepackage{hyperref}
\usepackage{xcolor}

\definecolor{linkcolor}{RGB}{0, 0, 255}      
\definecolor{citecolor}{RGB}{0, 128, 0}     
\definecolor{urlcolor}{RGB}{255, 0, 0}       

\hypersetup{
    colorlinks=true,
    linkcolor=linkcolor,
    citecolor=citecolor,
    urlcolor=urlcolor,
    linktoc=all,  
    pdfborder={0 0 0}  
}

\DeclareCaptionJustification{justified}{\leftskip=0pt \rightskip=0pt \parfillskip=0pt plus 1fil}

\begin{document}
\definecolor{dy}{rgb}{0.9,0.9,0.4}
\definecolor{dr}{rgb}{0.95,0.65,0.55}
\definecolor{db}{rgb}{0.5,0.8,0.9}
\definecolor{dg}{rgb}{0.2,0.9,0.6}
\definecolor{BrickRed}{rgb}{0.8,0.3,0.3}
\definecolor{Navy}{rgb}{0.2,0.2,0.6}
\definecolor{DarkGreen}{rgb}{0.1,0.4,0.1}

\title{Edge Spin fractionalization in open one-dimensional spin-$S$ quantum antiferromagnets}
\author{Pradip Kattel}
\email{pradip.kattel@rutgers.edu}
\affiliation{Department of Physics and Astronomy, Center for Material Theory, Rutgers University,
Piscataway, New Jersey, 08854, United States of America}
\author{Yicheng Tang}
\affiliation{Department of Physics and Astronomy, Center for Material Theory, Rutgers University,
Piscataway,  New Jersey, 08854, United States of America}
\author{J. H. Pixley}
\affiliation{Department of Physics and Astronomy, Center for Material Theory, Rutgers University,
Piscataway,  New Jersey, 08854, United States of America}
\affiliation{Center for Computational Quantum Physics, Flatiron Institute, 162 5th Avenue, New York, NY 10010}
\author{Natan Andrei}
\affiliation{Department of Physics and Astronomy, Center for Material Theory, Rutgers University,
Piscataway,  New Jersey, 08854, United States of America}

\begin{abstract}
We show that a gapped spin-$S$ chain  with antiferromagnetic (AFM) order exhibits in the thermodynamic limit  exponentially localized fractional $\pm \frac{S}{2}$ edge modes when the system possesses U(1) symmetry. We show this  for   integrable and non integrable spin chains both analytically and numerically. Through exact analytical solutions, we show that an AFM spin-$\frac{1}{2}$ chain with {\it explicitly} broken $\mathbb{Z}_2$ symmetry and an integrable AFM spin-$1$ chain with {\it spontaneously} broken  $\mathbb{Z}_2$ symmetry have  $\pm \frac{1}{4}$ and  $\pm \frac{1}{2}$ fractionalized edge modes, respectively. 
Furthermore, employing the density matrix renormalization group technique, we extend this analysis to {\it generic} $XXZ-S$ chains with $S\leq 3$ and  demonstrate that these fractional spins are robust quantum observables, substantiated by the observation of a variance of the associated fractional spin operators that is consistent with a vanishing functional form in the thermodynamic limit. Moreover, we find that the edge modes are robust to  disorder that couples to the N\'eel order parameter.
\end{abstract}

\maketitle

{\it Introduction}: The concept of quantum fractionalization, initially demonstrated by Jackiw and Rebbi in continuous quantum field theory \cite{PhysRevD.13.3398} and later extended to lattice systems by Su, Schrieffer, and Heeger \cite{PhysRevLett.42.1698}, has now been found in various other systems. Notable manifestations include spin-charge separation \cite{giamarchi2003quantum,tyner2023spin}, fractional spin excitations in spin liquids \cite{kane2005z,QSLbalents2010spin,QSLbroholm2020quantum,QSLsavary2016quantum,QSLzhou2017quantum, banerjee2016proximate,hermanns2018physics}, and the fractional quantum Hall effect \cite{laughlin1983anomalous,stormer1999fractional,stormer1999nobel,eisenstein1990fractional,bernevig2006quantum,kane2005quantum,roy2009topological}. It is well known theoretically that the exponentially localized fractional edge modes exist in systems with symmetry protected topological 
phases such as 
polyacetylene \cite{PhysRevLett.42.1698}, the spin-1 Haldane chain 
\cite{haldane1983continuum,haldane1983nonlinear,affleck1989quantum}, topological superconductors \cite{sato2017topological,kitaev2001unpaired,pasnoori2020edge,frolov2020topological,alicea2012new,leijnse2012introduction,wang2018weak,flensberg2021engineered}, and topological insulators \cite{moore2009next,bernevig2013topological,moore2007topological,wada2011localized,schnyder2009classification,qi2011topological,moore2010birth,hasan2011three,fu2007topological,ran2009one}. These topological edge modes and/or the fractional excitations have been experimentally observed in various natural and engineered systems including fractional excitation in spin-$\frac{1}{2}$ antiferromagnet \cite{lake2005quantum,kim2006distinct}, topological excitations in various systems  \cite{samajdar2023emergent,de2019observation}, topological edge modes in spin chains \cite{meier2016observation},  topological edge modes in a wide array of interacting and non-interacting topological systems \cite{zhang2018observation,kanungo2022realizing,yan2019topological,barik2016two,kim2014edge,chen2009experimental,gong2019experimental}.

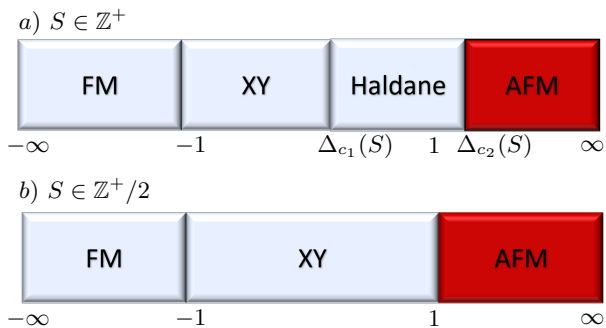
\begin{figure}[htb!]
    \centering
    \includegraphics[width=0.9\linewidth]{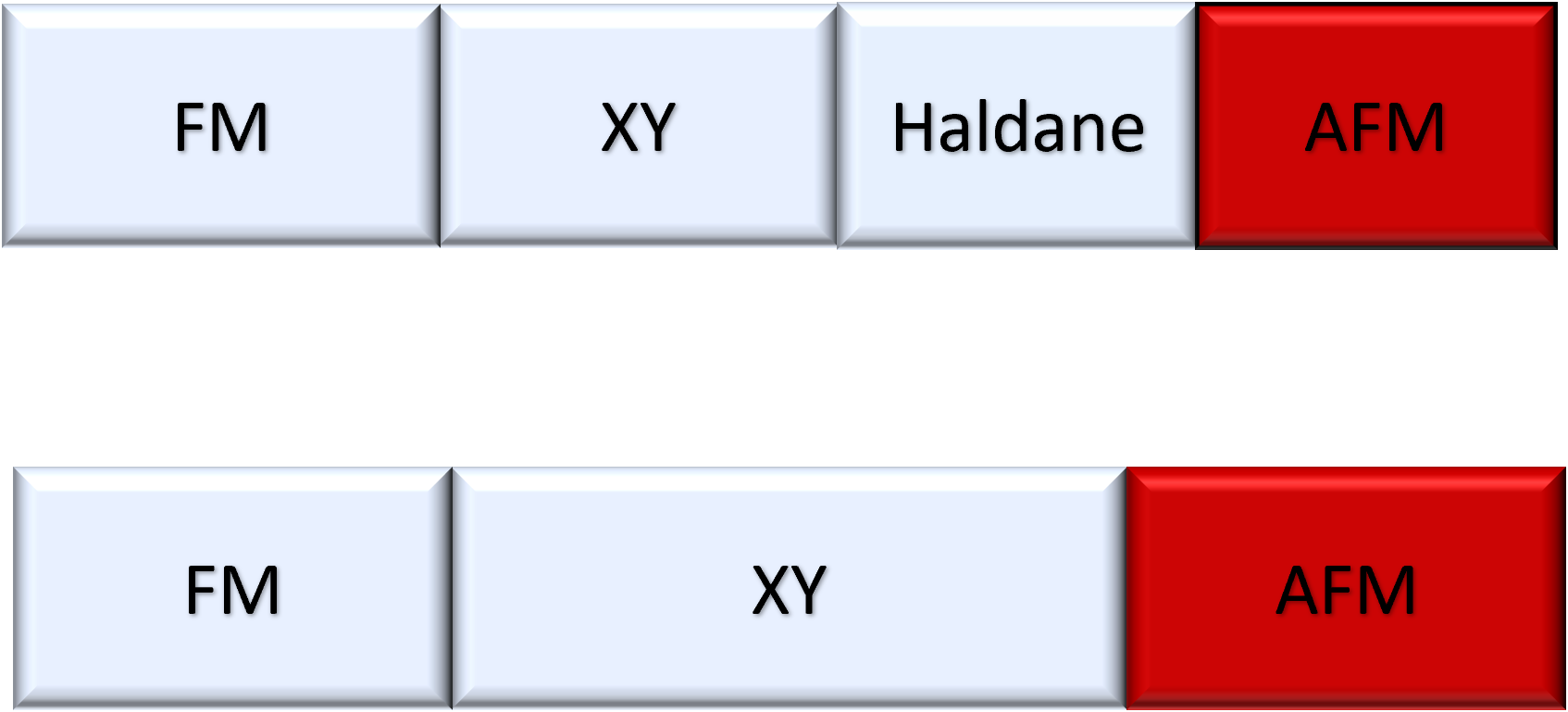} 
    \begin{tikzpicture}[overlay, remember picture]
         \node[black] at (-7.75,2.1) {$-\infty$};
        \node[black] at (-5.59,2.1) {$-1$};
        \node[black] at (-3.4,2.1) {$\Delta_{c_1}(S)$};
        \node[black] at (-1.55,2.1) {$\Delta_{c_2}(S)$};
        \node[black] at (-2.35,2.1) {$1$};
        \node[black] at (-0.25,2.1) {$\infty$};
        \node[black] at (-7.15,3.75) {$a)~S\in\mathbb{Z}^+$};
        \node[black] at (-7.00,1.5) 
        {$b)~S\in\mathbb{Z}^+/2$};
        \node[black] at (-7.75,-0.2) {$-\infty$};
        \node[black] at (-5.59,-0.2) {$-1$};
        \node[black] at (-2.35,-0.2) {$1$};
        \node[black] at (-0.25,-0.2) {$\infty$};
    \end{tikzpicture}
    \hspace{0.75cm}
    \caption{{\bf Phase diagrams of the $XXZ-S$ chains} a) for integer spin  and b) half-integer spin chain. In the integer chain there is a gapped topological phase between $\Delta_{c_1}<\Delta<\Delta_{c_2}>1$ 
    between the gapless XY and gapped antiferromagnetic phase whereas in the half-integer chain there is a direct BKT transition from gapless XY phase to the gapped antiferromagnetic phase. As  integer $S$ increases,  $\Delta_{c_2}\to 1^+$ and $\Delta_{c_1}\to 1^-$  such that the value of $\Delta_{c_2}-\Delta_{c_1}$ decreases. Here we are only interested in the antiferromagnetic phase highlighted in the red color in each of the phase diagrams.    }
    \label{fig:phasediag}
\end{figure}

In this letter we explicitly show  that exponentially localized edge modes appear in many other contexts, not associated with topology but rather with symmetry breaking  and  gapfulness,  extending   to a  broad  context results
obtained  for the integrable XXZ model~\cite{pasnoori2023spin}. In particular, we consider the spin-$S$ quantum antiferromagnetic chain and we make the hypothesis that exponentially localized fractional spin $\pm \frac{S}{2}$ edge  modes appear if the model satisfies (at least) the following three properties:
\begin{enumerate}
    \item Antiferromagnetic order in the bulk due to explicitly or spontaneously broken $\mathbb Z_2$ symmetry in the thermodynamic limit,
    \item  $U(1)$ symmetry,
    \item A finite bulk gap. 
\end{enumerate}
These conditions allow models that are integrable or not, nearest neighbor or not. In the following, we mainly consider the quantum Heisenberg spin-$S$ anisotropic chain defined by the Hamiltonian
\begin{equation}
   H_\Delta= \sum_{i=1}^{N-1}(\vec S_i\cdot\vec S_{i+1})_\Delta=\sum_{i=1}^{N-1} S^x_i S^x_{i+1}+S^y_i S^y_{i+1}+\Delta S^z_i S^z_{i+1}.
    \label{ham}
\end{equation} 
where $S^x_i, S^y_i$ and $S^z_i$ are the spin-$S$ operators  that belong to the spin-$S$ representation of $SU(2)$ at site $i$. The model is non-integrable  for $S>1/2$. We will also consider several perturbations to this model that either satisfy or violate the three conditions defined in the introduction. The phase diagram of these models is well known \cite{kjall2013phase} and is shown in Fig.~\ref{fig:phasediag}. Our hypothesis  applies to  the antiferromagnetic regime \textit{i.e.} $\Delta>1$ for half-integer $S$ and $\Delta>\Delta_{c_2}$ for integer $S$,  where $\Delta_{c_2}>1$ is the boundary between the gapped Haldane phase and the gapped antiferromagnetic phase.  
 It was recently  established that $XXZ-\frac{1}{2}$ chain  in the gapped antiferromagnetic regime hosts fractionalized $\pm\frac{1}{4}$ edge spins~\cite{pasnoori2023spin}. When projected onto the low energy subspace spanned by the ground state these can be
identified with the strong zero energy mode discussed in \cite{fendley2016strong, fendley2012parafermionic,yates2020dynamics,yates2021strong,vasiloiu2019strong,zvyagin2021majorana,zvyagin2022charging,zvyagin2024strong}.

Here we generalize the construction to arbitrary spin$-S$ chains and show that fractional $\pm S/2$ spin modes appear at the edges of the open chain. We demonstrate that  these spin modes correspond to  genuine local
quantum observables as the variance of the edge spin operators vanish in the thermodynamic limit. These conclusions  are also robust to disorder that couples to the N\'eel order parameter.

 We begin by considering a solvable spin $S=1/2$  model in a staggered magnetic field. It  provides an example of a system where the appearance of  $S=\pm 1/4$ is driven by explicit symmetry breaking. We then study an integrable $S=1$
model  and show  the presence of fractionalized edge modes of spin $\pm 1/2$ using the Bethe Ansatz \cite{bethe1931theorie,faddeev1996algebraic,sutherland2005introduction}. We then use DMRG to further verify these edge modes are robust in non-integrable  $XXZ-S $ spin chains $H_\Delta$  up to spin $S=3$.

 We begin by examining the analytically tractable spin-1/2 $XX$ open chain with a staggered magnetic field
\begin{equation}
  H_h=\sum_{i=1}^{N-1}(-1)^i h S^z_i+ (\vec S_i\cdot\vec S_{i+1})_{\Delta=0}. 
\end{equation}
The staggered magnetic field opens a gap in the model, leading to the development of antiferromagnetic order from the explicit breaking of spin flip $\mathbb{Z}_2$ symmetry. Since this model can be mapped to free fermions via the Jordan-Wigner transformation, we can compute the spin profile $S^z_j$ exactly and establish that it fractionalizes in the ground state to $\pm\frac{1}{4}$ at the two edges (see supplement for details). 


Consider now the integrable spin $S=1$ open chain  \cite{yang2006q,frappat2007complete}  given by the  Hamiltonian
\begin{equation}
    H_1=\sum_{i=1}^{N-1}(\vec S_i\cdot\vec S_{i+1})_\Delta + H^I
\end{equation}
where $H^I$ is the integrable deformation whose explicit form is given in Eq.(S.34) supplementary material.

We solve this model using the Bethe Ansatz \cite{cherednik1984factorizing,sklyanin1988boundary,fan1996algebraic,wang2015off}.
and show that the model possesses fractionalized $ S=\pm \frac{1}{2}$ edge modes by carefully studying the role of the boundary string solutions.  The root distribution before adding the boundary string has total spin $1$ for even number of sites $N$ which upon adding the $0$ energy boundary string solution gives the ground state configuration with the total spin $S^z=0$. This indicates that the ground state possess fractional $\pm\frac{1}{2}$ spin (see supplementary material for details). Further, these modes are stable against  perturbations in the bulk that respects the  three conditions in the introduction. This is expected since the correlations are exponentially decaying  so that the boundary physics can not be effected. Thus, we expect that also the $XXZ-1$ chain without the $H^I$ terms  would exhibit this property for $\Delta>\Delta_{c_2}$ where antiferromagnetic order develops. We  explicitly verify this claim below  using DMRG where we can  access  the spin profile $S^z_j$.

 We may continue to construct  integrable models with higher spin-$S$ using the fusion technique \cite{wang2015off}, and  show that these models have $\pm \frac{S}{2}$ edge spin accumulation described by various boundary strings just like in the spin-$\frac{1}{2}$ and spin-$1$.   The  Hamiltonians for these integrable models are fine tuned and  have various higher order (cubic, quartic etc) spin-spin interactions required for integrability. These terms however  can be dropped, as we argued above, in the $\Delta>\Delta_{c_2}$ regime when studying the edge spin fractionalization. 
 
 We now turn to verify these claims by studying spin chains numerically.  By means of DMRG  we  show  that    $XXZ-S$ spin chains  given by Eq.~\eqref{ham}  possess   $\pm \frac{S}{2}$ edge modes  that are sharply localized, and furthermore,  they are robust against  any perturbation satisfying the three conditions in the introduction. We also show, following arguments developed in  \cite{pasnoori2023spin}, that the operators associated with these edge spin accumulations have 
a variance whose functional form vanishes in the thermodynamic limit. 
This numerically demonstrates that these edge spins are not mere  quantum averages but rather sharp quantum observables. We verify this claim by computing the spin profile for various values of anisotropy parameters $\Delta$ and system size $N$  for both half-integer and integer spin-$S $ chains for $1/2\leq S\leq 3$. All of our DMRG calculation are performed by using ITensors library \cite{itensor} by setting truncation cut-off of the singular values at $10^{-10}$. The spin$-\frac{1}{2}$ results were first obtained in \cite{pasnoori2023spin}.

\begin{figure*}[ht!]
    \begin{subfigure}[b]{0.3\textwidth}        \includegraphics[width=\textwidth]{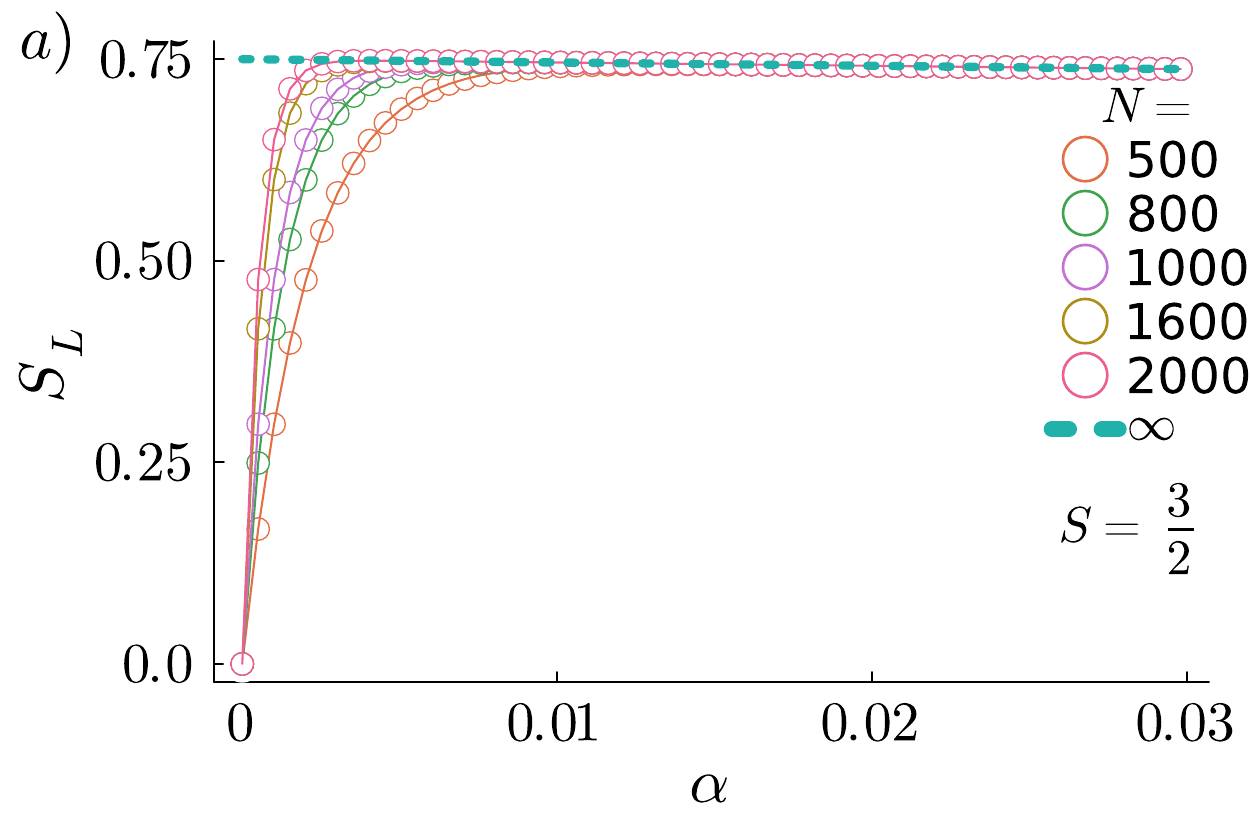}
    \end{subfigure}
    \begin{subfigure}[b]{0.3\textwidth}
\includegraphics[width=\textwidth]{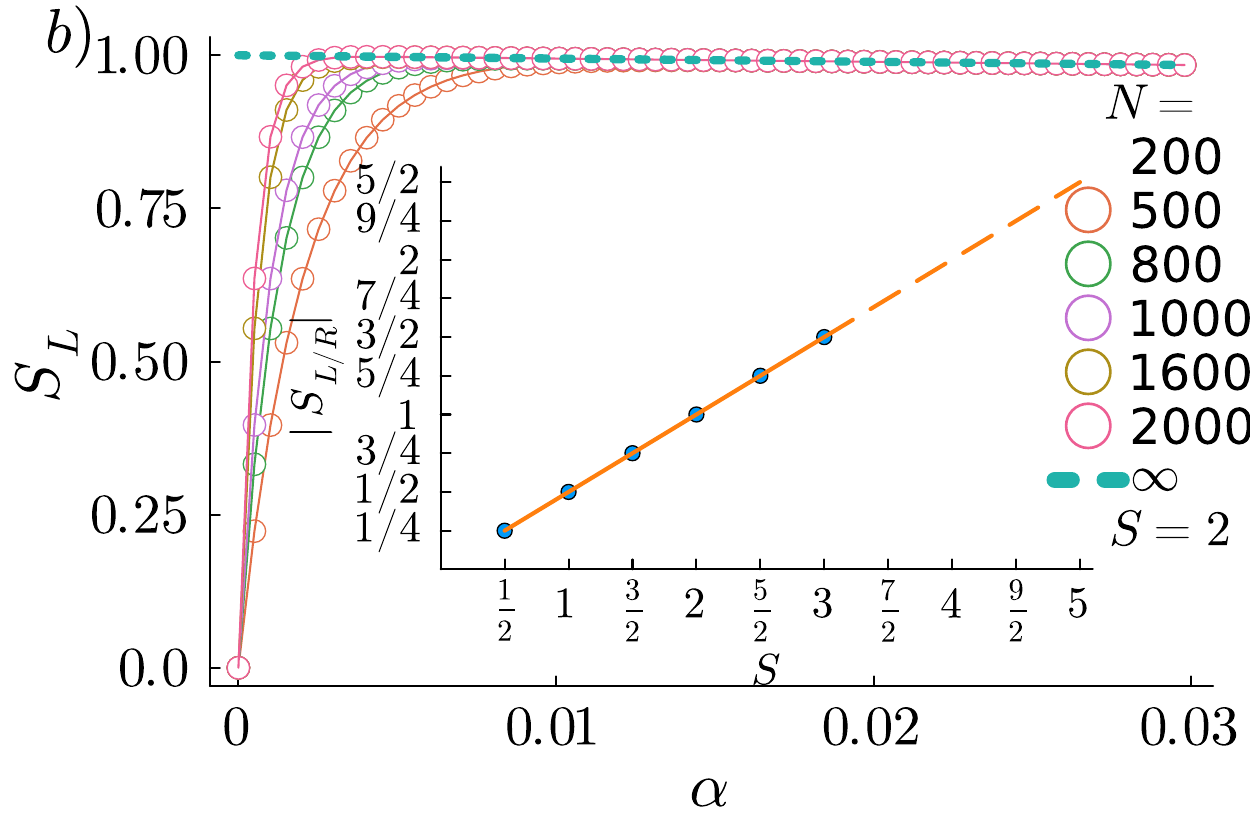}
    \end{subfigure}
    \begin{subfigure}[b]{0.3\textwidth}
\includegraphics[width=\textwidth]{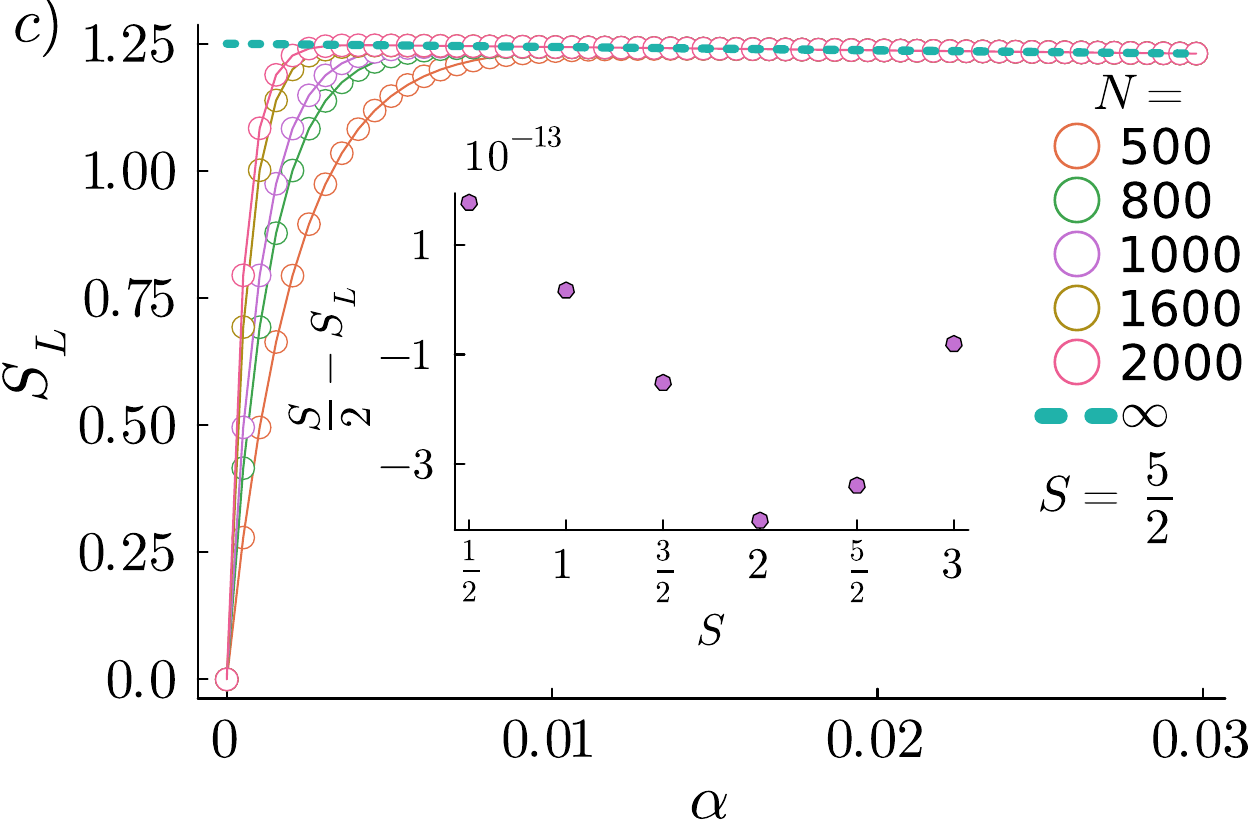}
    \end{subfigure}
        \begin{subfigure}[b]{0.3\textwidth}
        \includegraphics[width=\textwidth]{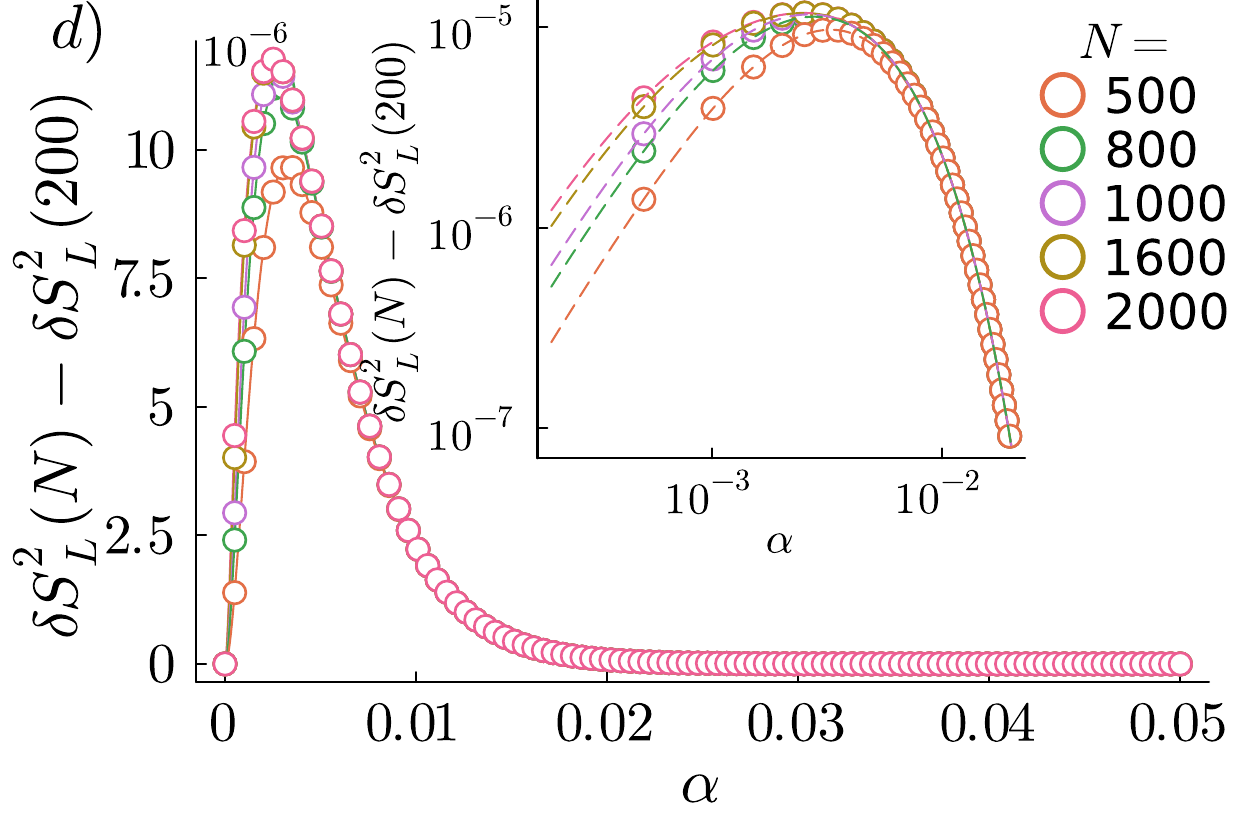}
    \end{subfigure}
    \begin{subfigure}[b]{0.3\textwidth}
        \includegraphics[width=\textwidth]{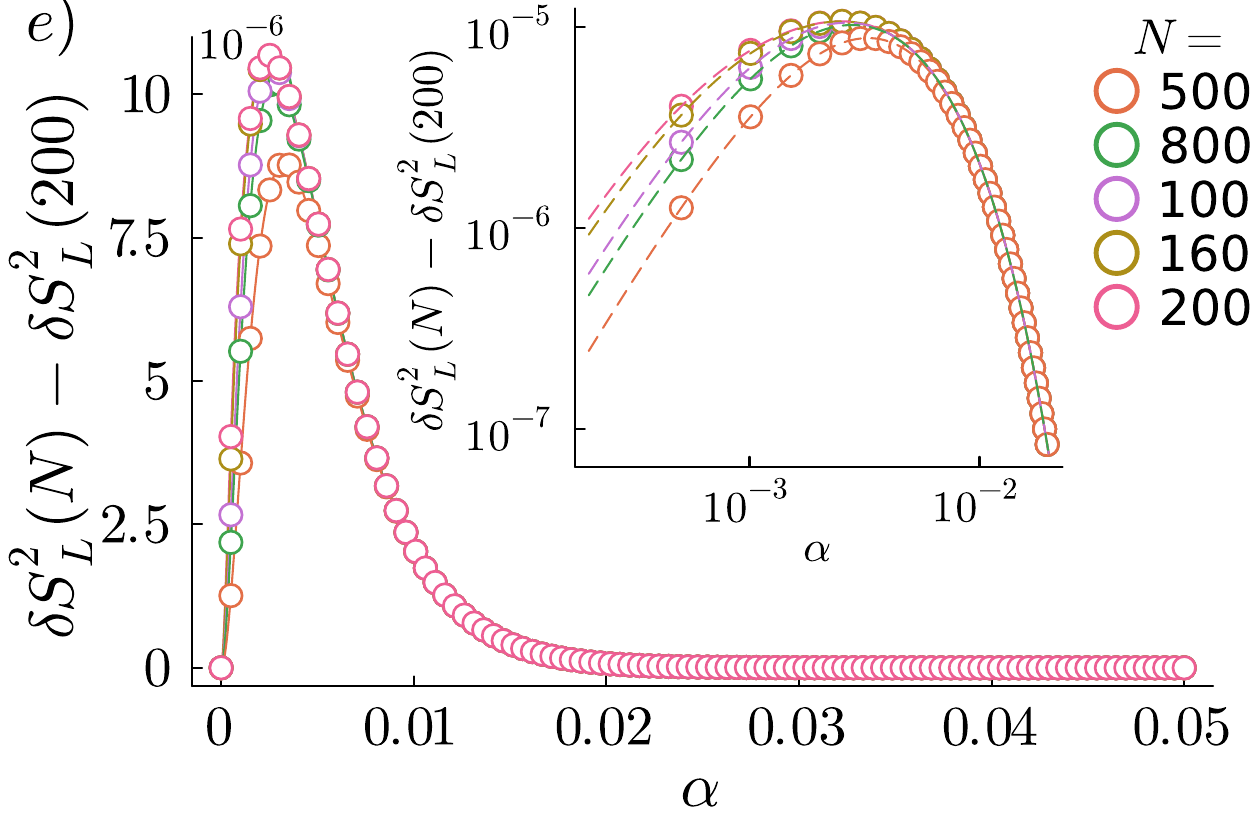}
    \end{subfigure}
    \begin{subfigure}[b]{0.3\textwidth}
        \includegraphics[width=\textwidth]{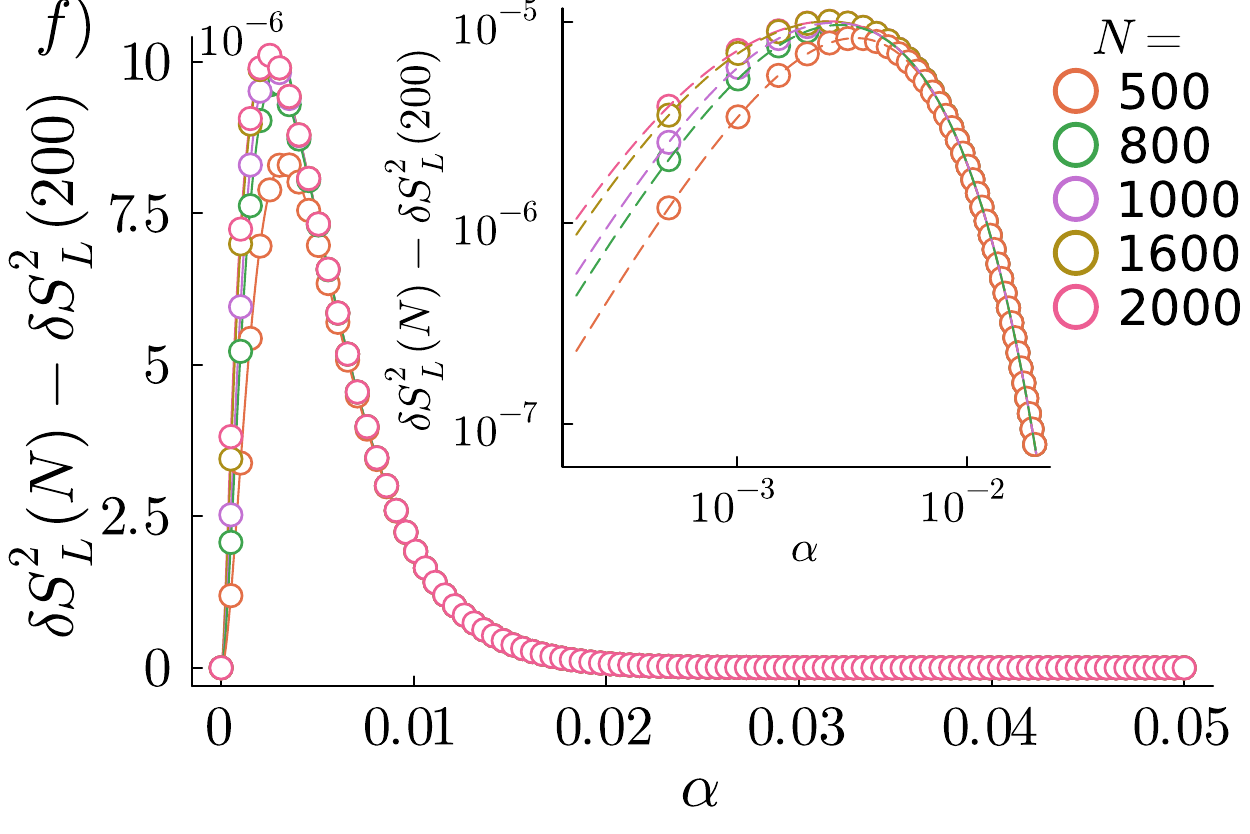}
    \end{subfigure}
    \caption{
    {\bf Spin accumulation and variance up to $S=3$}:
    The finite size scaling of the left localized a) $ \frac{3}{4}$ spin for $S=\frac{3}{2}$ and $\Delta=2$ $XXZ$ chain b) $1$ spin for $S=2$ and $\Delta=2$ $XXZ$ chain. The middle inset shows the DMRG result for $S\leq 3$ and extrapolation to higher spin by fitting the data c) $ \frac{5}{4}$ spin for $S=\frac{5}{2}$ and $\Delta=2$ $XXZ$ chain. The right inset shows the difference between the expected outcome $\frac{S}{2}$ and DMRG result. The result is within the accuracy of the DMRG calculation. d) The variance of the edge spin operators $\delta S^2_{L}(N,\alpha)$ as well as their fit with the Ansatz in Eq.~\eqref{ansatzvar} are shown in d), e) and f) for the edge spins shown in a), b) and c)respectively. The insets shows the fit with the Ansatz for small values of $\alpha$.
    }
\label{explocalization}
\end{figure*}

Following \cite{jackiw1983fluctuations,kivelson1982fractional,pasnoori2023spin}, we define the fractional edge spin operators
\begin{align}
     \hat S^z_{L}&=\lim_{\alpha\to 0}\lim_{N\to \infty}\hat S^z_{L}(N,\alpha)=\lim_{\alpha\to 0}\lim_{N\to \infty} \sum_{j=1}^N e^{-\alpha j} S^z_j,
     \label{leftspop}
\end{align}
and $\hat S^z_R$ is defined similarly but instead with the decay factor $\exp\{-\alpha(N+1 -j)\}$. These operators have quantized expectation values $\pm \frac{S}{2}$ in the ground state.   We show their variance $\delta \hat S^2_{L/R}$ vanishes in the ground state, which indicates the fractional operator is a sharp quantum observable. In order to compute the variance numerically, we follow \cite{pasnoori2023spin} and propose that the variance satisfy following Ansatz in the thermodynamic limit
 \begin{equation}
     \delta S^2_{L/R}(N,\alpha)= \delta S^2_{L/R}(\infty,\alpha)-A \alpha e^{-B \alpha N}.
     \label{ansatzvar}
 \end{equation}
With the help of this Ansatz, we can  compute the variance in the thermodynamic limit as $\delta S^2_{L/R}=\lim_{\alpha\to 0} \delta S^2_{L/R}(\infty,\alpha)$. We verified the Ansatz for various values of $N$ for Hamiltonian Eq.~\eqref{ham} as shown in the representative cases in Fig.~\ref{explocalization} as it fits the data quite well.

We compute the edge spin accumulation and the variance of the operators for Hamiltonian Eq.~\eqref{ham} for $S=\{\frac12,1,\frac32,2,\frac52,3 \}$ and show that the fractional spin of magnitude $\frac{S}{2}$ exist in the edges and the variance of the spin operators vanish in each of these cases.  The magnitude of localized edge modes for these cases are shown in Fig.~\ref{explocalization} and more plots for individual $S$ values are shown in the supplementary materials.  In addition, more cases of perturbations that respect the three conditions and hence possessing the sharply localized edge modes are shown in the supplementary materials.

\begin{figure}[htbp]
  \centering
  \begin{subfigure}{0.49\columnwidth}
    \centering
\includegraphics[width=\columnwidth]{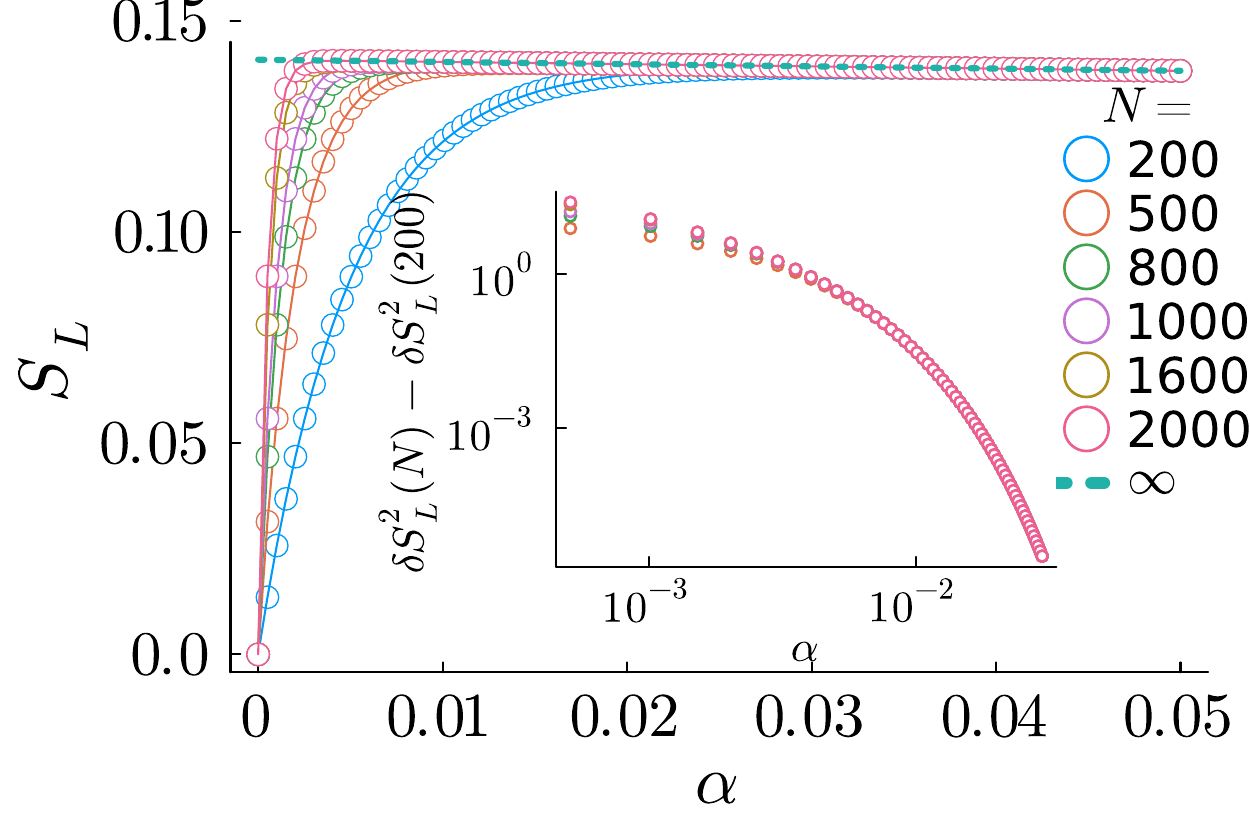}
  \end{subfigure}
  \begin{subfigure}{0.49\columnwidth}
    \centering
    \includegraphics[width=\columnwidth]{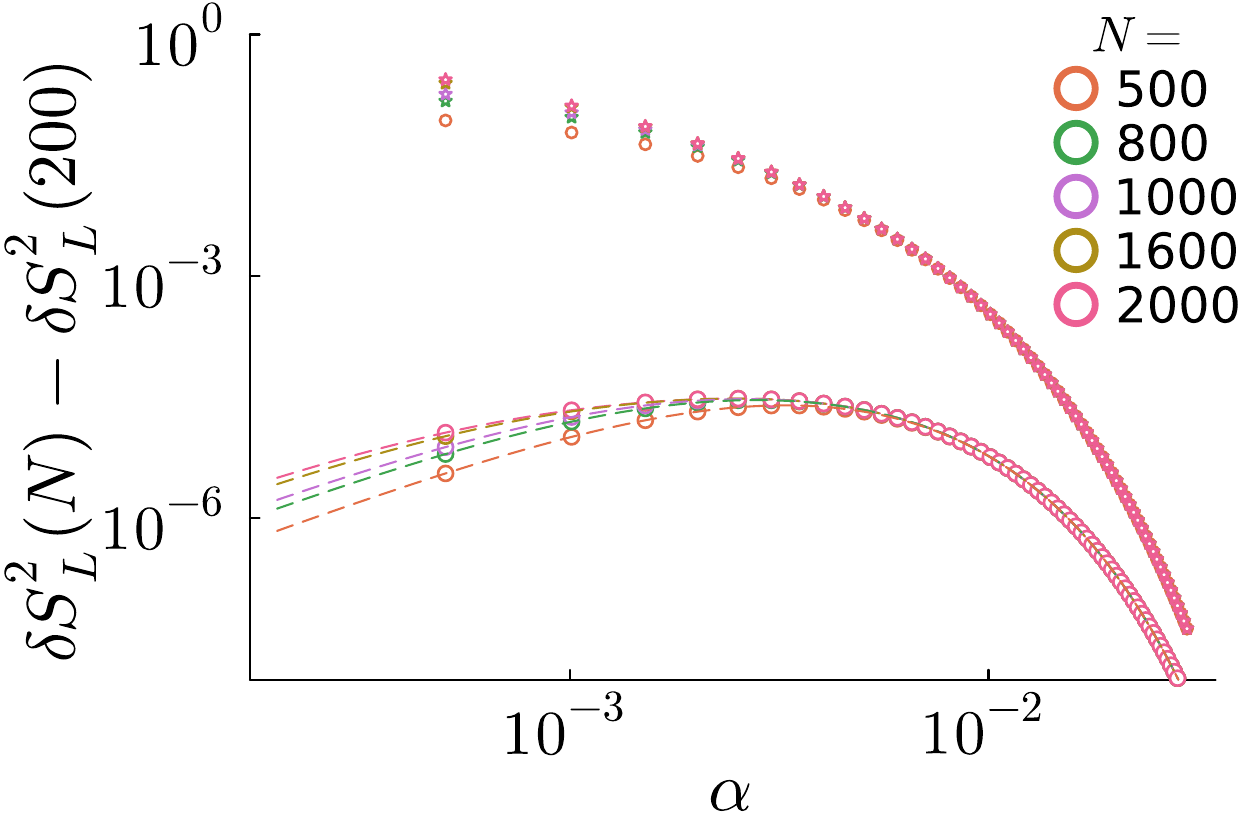}
  \end{subfigure}
  \caption{{\bf Breaking U(1) symmetry}: The left panel shows that edge spin accumulation in the antiferromagnetic regime of transverse field Ising mode is not quantized to $\frac{S}{2}$. The inset shows that the variance of the edge spin operator  does not vanish in the thermodynamic limit. The right figure shows that the variance of the left edge spin operator the spin-1 XYZ model (shown in circular marker) with couplings  $J_x=1, J_y=0.85$ and $J_z=1.5$ does not vanish which is consistent with the fact that it does not satisfy the three conditions (shown with star marker) and  the vanishing variance of the left edge spin operator the spin-1 $XXZ$ model with couplings  $J_x=J_y=1$ and $J_z=1.5$. The variance is fitted with the Ansatz Eq.~\eqref{ansatzvar} for the latter case.}
\label{xxz-xyz-comp}
\end{figure}

{\it Breaking U(1) symmetry}:
To complete the argument we present some  cases where the conditions outlined above are not satisfied and show that as a result the edge modes are no longer quantized. Consider the transverse field Ising model
\begin{equation}
    H_{\mathrm{TFI}}=\sum_{i=1}^{N-1} \sigma_i^z\sigma_{i+1}^z + g \sum_{i=1}^N \sigma_i^x
\end{equation}
in the antiferromagnetic phase. Since there is no $U(1)$ conservation, the criteria is violated. In such case, the spin accumulation is not quantized and the variance of the edge spin operators do not vanish as shown in Fig.~\ref{xxz-xyz-comp} for the representative case of $g=0.75$ for the spin accumulation in the left edge. A similar situation occurs in the XYZ model
\begin{equation}
    H_{\mathrm{XYZ}}=\sum_{i=1}^{N-1} J_x S^x_i S^x_{i+1}+J_y S^y_i S^y_{i+1}+ J_z S^z_i S^z_{i+1}
\end{equation}
as shown in bottom panel of Fig.~\ref{xxz-xyz-comp} that as soon as the $U(1)$ symmetry is broken, the variance of the edge spin operator ceases to vanish.  Let us start from the $XXZ-1$ with the couplings $J_x=J_y=1$ and $J_z=1.5$ where the edge spin is quantized to $\pm \frac{1}{2}$ and the variance of the edge spin vanishes. Now, we break the $U(1)$ symmetry by choosing the couplings $J_x=1, J_y=0.85$ and $J_z=1.5$. In this case, the variance of the edge spin operators do not vanish as shown in Fig.~\ref{xxz-xyz-comp}. This shows that the $U(1)$ symmetry associated with the conservation of the $z-$component of the spin is essential for the existence of the robust edge spin. A detailed understanding of this effect we leave for future work.

{\it Breaking Translational Symmetry by Disorder}: We break the translational symmetry in the system by adding disorder of the form $H_D=\sum_j  (-1)^j W_j S^z_j$, where $W_j$ are  random positive fields uniformly chosen from $[0,W]$, and the full Hamiltonian is now $H=H_{\Delta}+H_D$. This form of the disorder is chosen because it preserves the antiferromagnetic order in the bulk on average thereby satisfying all of our three criteria. We find that the edge mode is robust to this kind of disorder where not only the edge mode is quantized to $\pm \frac{S}{2}$ but also the variance of the edge spin operator vanishes in the thermodynamic limit. The effects of gap closing disorder are left for future work as they are more subtle because it  induces spinon exciations in the bulk and it is unclear, at present, how to analyze their contribution to the variance.

{\it General Argument}: Having provided all these examples of our hypothesis, we turn to discuss the argument for the formation of these $\pm S/2$ edge modes. The argument is based on the observation  that { the low energy physics of} a generic $XXZ-S$ chain can be { described by} $2S$ copies of spin-$\frac{1}{2}$ chains with specific couplings between the chains \cite{schulz1986phase,giamarchi2003quantum}, where the spin $S$ operator in each site can be formally written as $ S_i=\sum_{n=1}^{2S}\sigma_{n,i}$, subject to constraints,  with  $\sigma_{n,i}$ being the spin-$\frac{1}{2}$ spin operators with $i$ being the site index and $n$  the chain index. Then, provided that these  additional couplings satisfy the three conditions in the introduction, we expect that there exists $2S\times \left(\pm \frac{1}{4}\right)=\pm \frac{S}{2}$ edge modes which is a result due to $\pm \frac{1}{4}$ edge modes contribution from each of the $2S$ copies of spin $1/2$ chains. This is a natural presumption as models with a gap have exponentially decaying correlation functions such that the boundary physics can not be effected significantly by bulk perturbations that do not close the energy gap and the inter-layer couplings are irrelevant deep in the antiferromagnetic phase. See supplementary materials for concrete construction for $S=1$ chain.

{\it Conclusion:} we found that in the ground state of spin$-S$ spin chains in the gapped antiferromagnetic phase, there are fractional spin operators at the edges with eigenvalues $\pm \frac{S}{2}$. We presented two models where this argument was demonstrated exactly and further verified our argument for $XXZ-S$ chains up to spin $S=3$ by means of DMRG. While we have concentrated on $XXZ-S$ Hamiltonians we hypothesize that actually any open spin-$S$  chain that satisfies the three conditions stated above will possess $\frac{S}{2}$ robust edge excitations. Extensions to SU(N) magnets and systems with discrete symmetries are left to future work.

Acknowledgment: We are grateful to T. Giamarchi for valuable suggestions and to P. Azaria and P. Pasnoori for insightful discussions and collaborations on related works. J.H.P. is partially supported by NSF Career Grant No.~DMR- 1941569.

\bibliography{ref .bib}

\widetext
\newpage
\begin{center}
    \textbf{\large Supplementary information of `Edge Spin fractionalization in one-dimensional spin-$S$ quantum antiferromagnets'}\\
    \vspace{0.4cm}
    Pradip Kattel,$^1$ Yicheng Tang,$^1$ J.~H.~Pixley,$^{1,2}$ and Natan Andrei$^1$\\
    \vspace{0.2cm}
    \small{$^1$\textit{Department of Physics and Astronomy, Center for Material Theory,
Rutgers University, Piscataway, New Jersey, 08854, United States of America}}\\
\small{$^2$\textit{Center for Computational Quantum Physics, Flatiron Institute, 162 5th Avenue, New York, NY 10010}}
\end{center}

\setcounter{equation}{0} 
\setcounter{page}{1}

\renewcommand{\theequation}{S.\arabic{equation}}

In this supplementary material, we provide additional data and detailed analysis to support our main findings on `Edge spin fractionalization in one-dimensional spin-$S$ quantum antiferromagnets'. First, we present additional DMRG results for larger values of spin $S$ and demonstrate spin accumulation in two-fold degenerate ground states with a representative example in Sec.\ref{sec:spin-profile}. We then explore various perturbations to the $XXZ-S$ model that respect the three conditions outlined in the main text, showing that these perturbations result in quantized edge spin accumulation and that the variance of the edge spin operator vanishes. Finally, we offer detailed analytical computations of the edge spin accumulation for three different models: the XX-$\frac{1}{2}$ model with a staggered magnetic field using the free-fermion technique in Sect.\ref{sec:freefermion}, and the integrable deformation of the $XXZ-1$ chain using the Bethe Ansatz in Sec.\ref{sec:spin1}.

\section{Edge spin accumulation}\label{sec:spin-profile}
\begin{figure}[H]
  \centering
  \begin{subfigure}[b]{0.3\textwidth}
    \includegraphics[width=\textwidth]{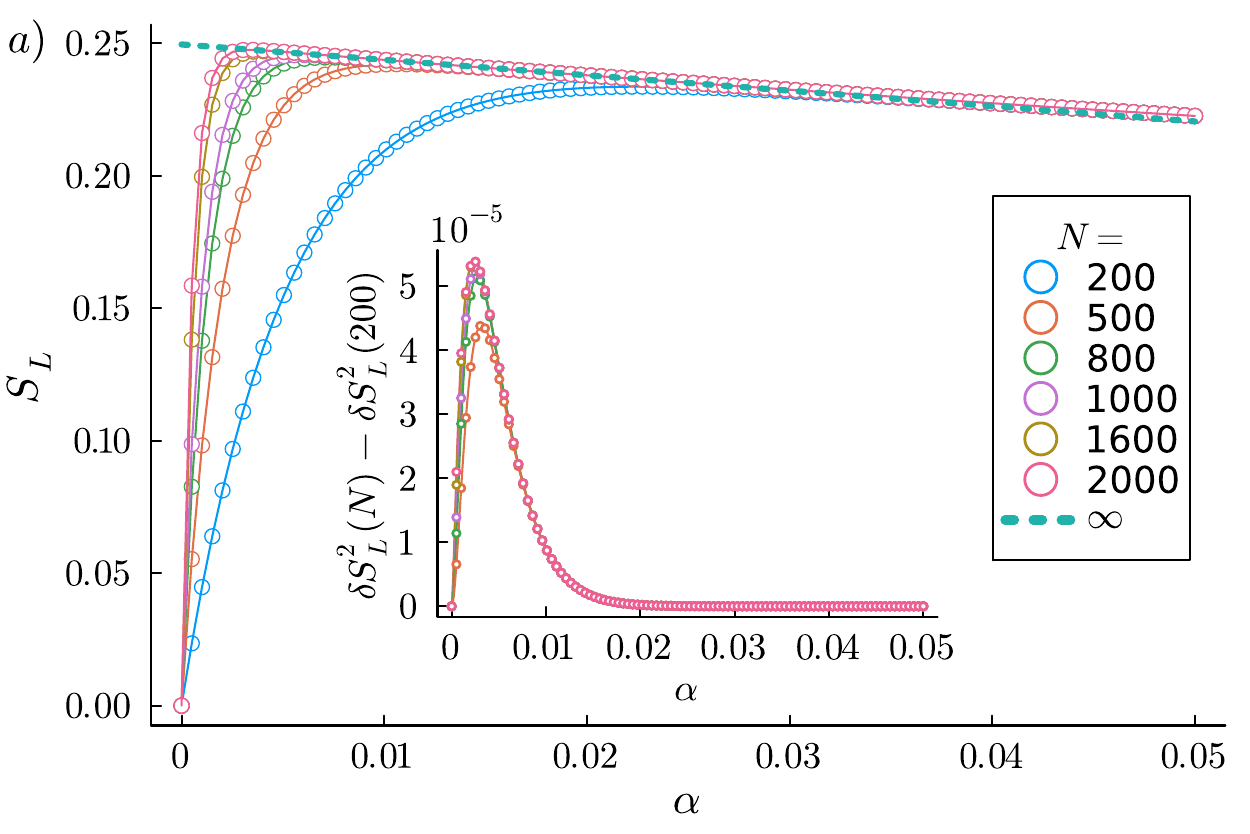}
    \label{fig:fig1}
  \end{subfigure}
  \hfill
  \begin{subfigure}[b]{0.3\textwidth}
    \includegraphics[width=\textwidth]{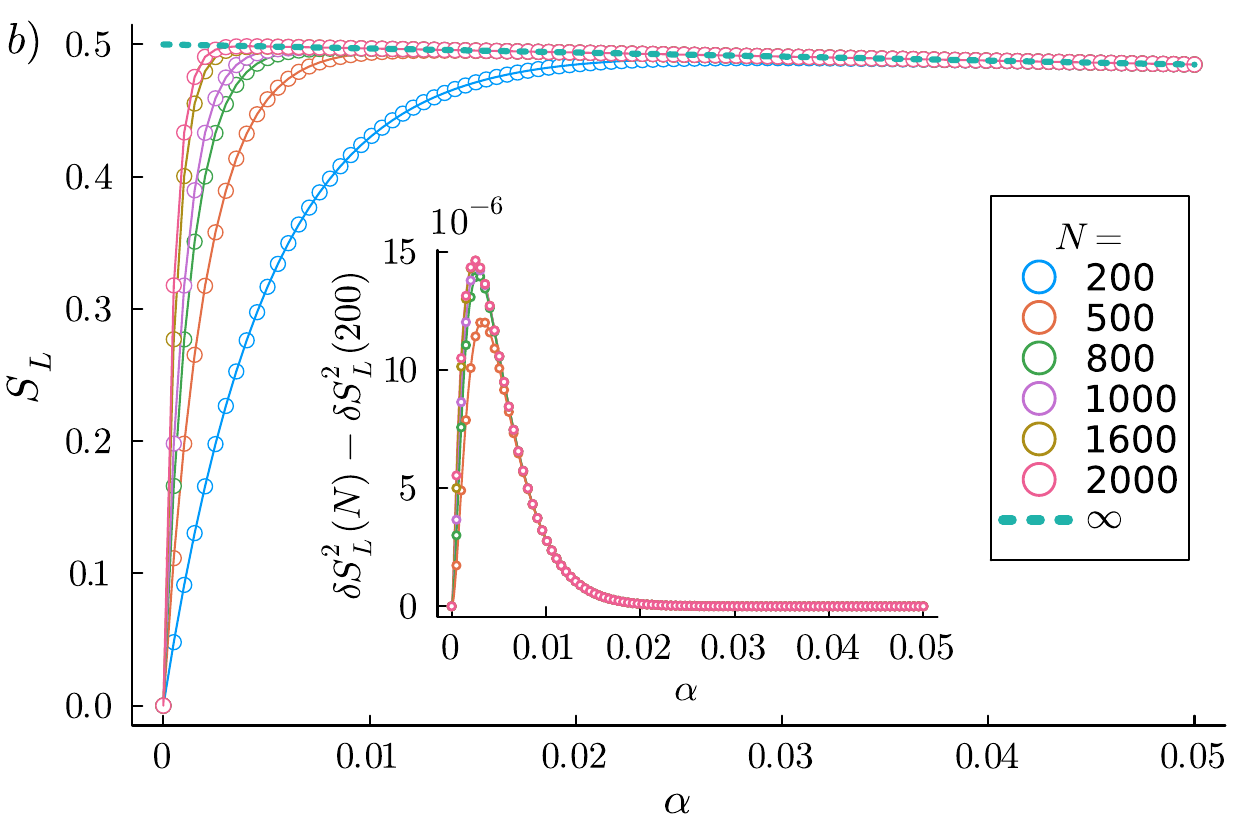}
    \label{fig:fig2}
  \end{subfigure}
  \hfill
  \begin{subfigure}[b]{0.3\textwidth}
    \includegraphics[width=\textwidth]{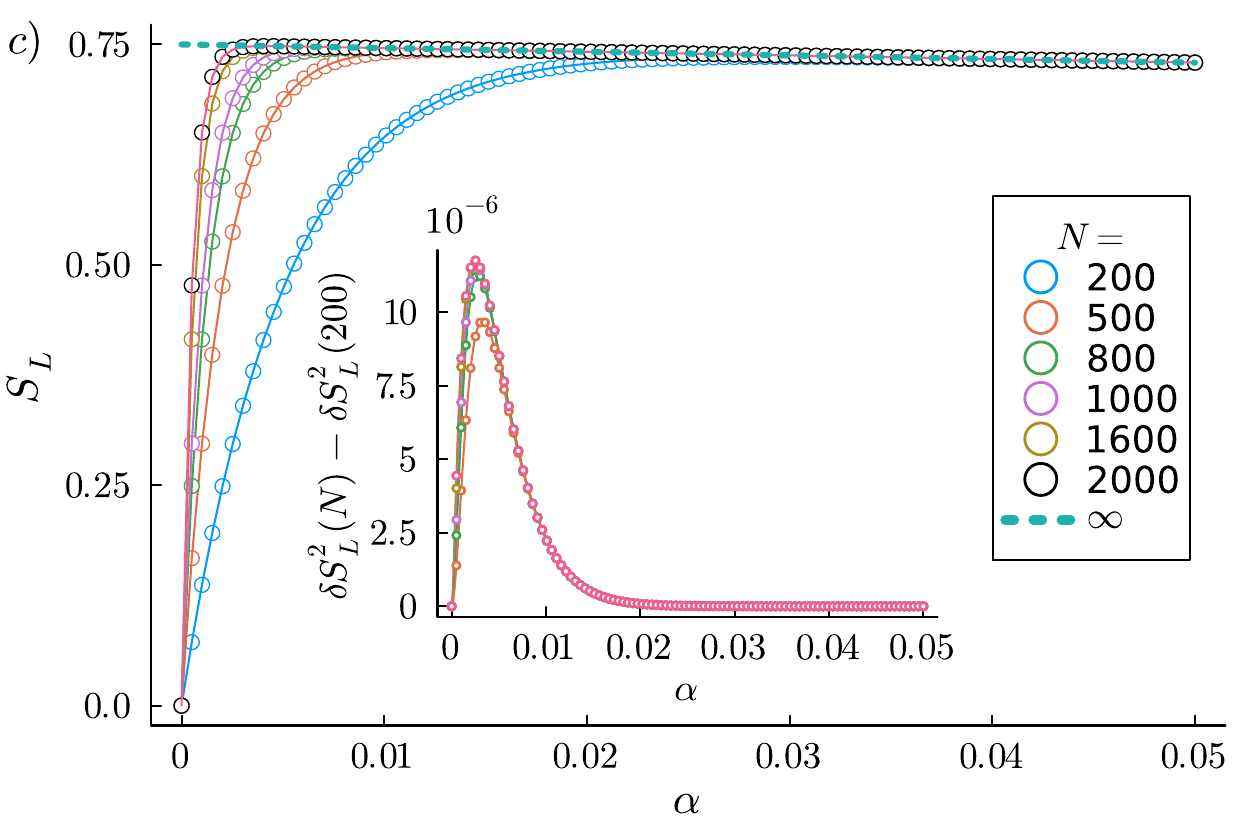}
    \label{fig:fig3}
  \end{subfigure}

  \medskip

  \begin{subfigure}[b]{0.3\textwidth}
    \includegraphics[width=\textwidth]{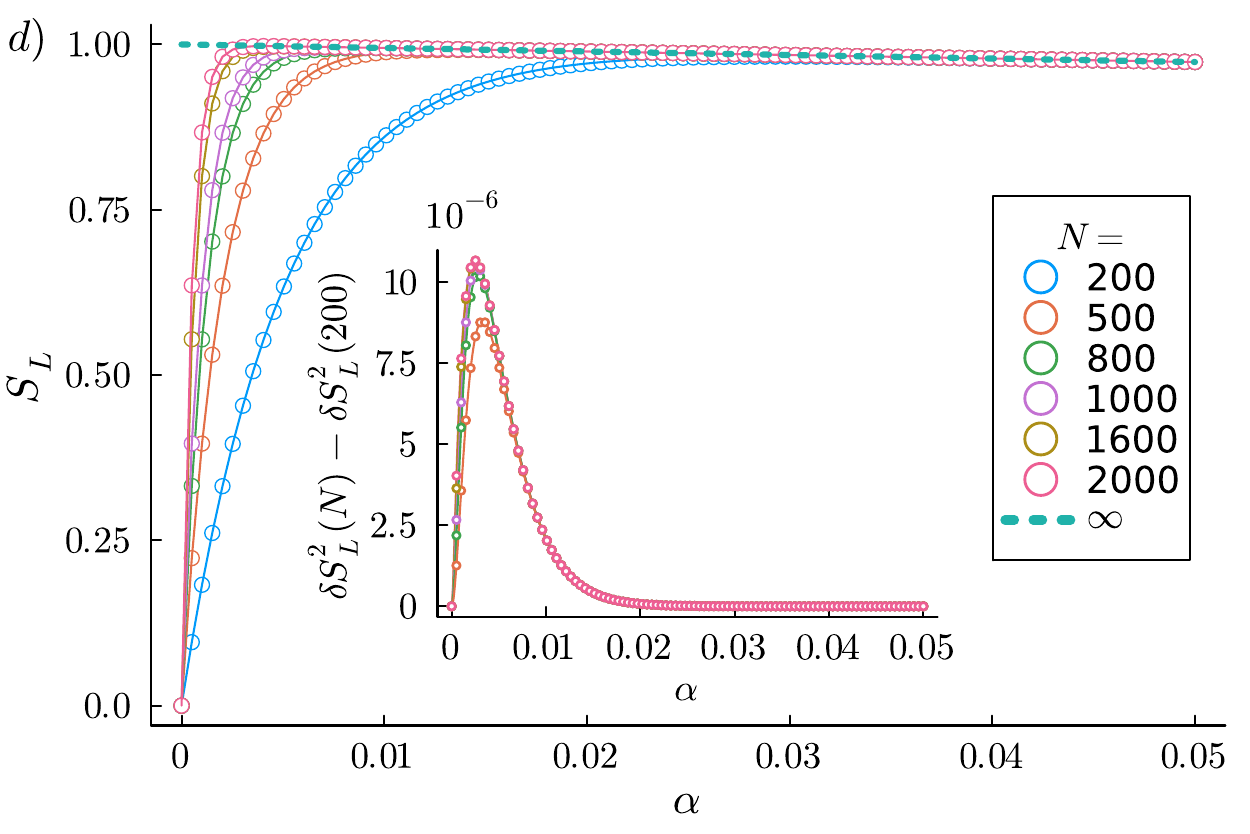}
    \label{fig:fig4}
  \end{subfigure}
  \hfill
  \begin{subfigure}[b]{0.3\textwidth}
    \includegraphics[width=\textwidth]{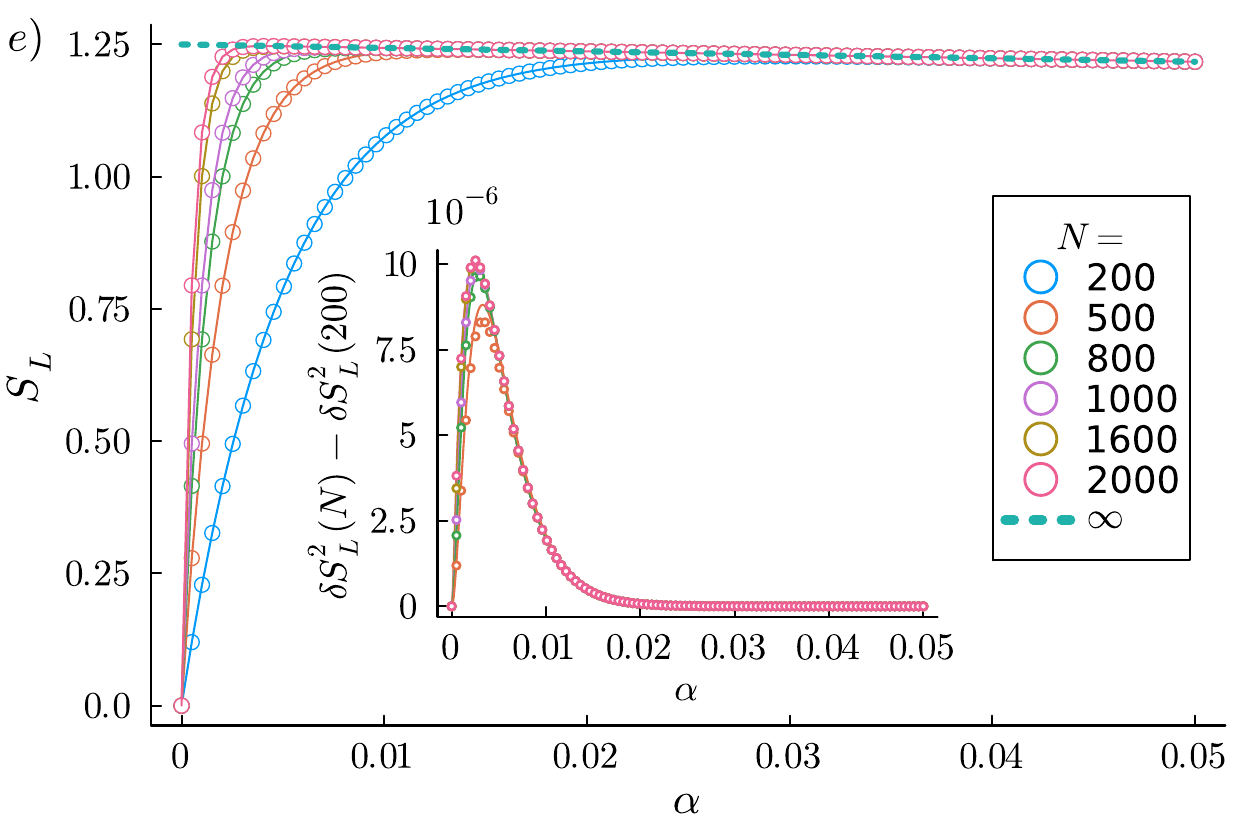}
    \label{fig:fig5}
  \end{subfigure}
  \hfill
  \begin{subfigure}[b]{0.3\textwidth}
    \includegraphics[width=\textwidth]{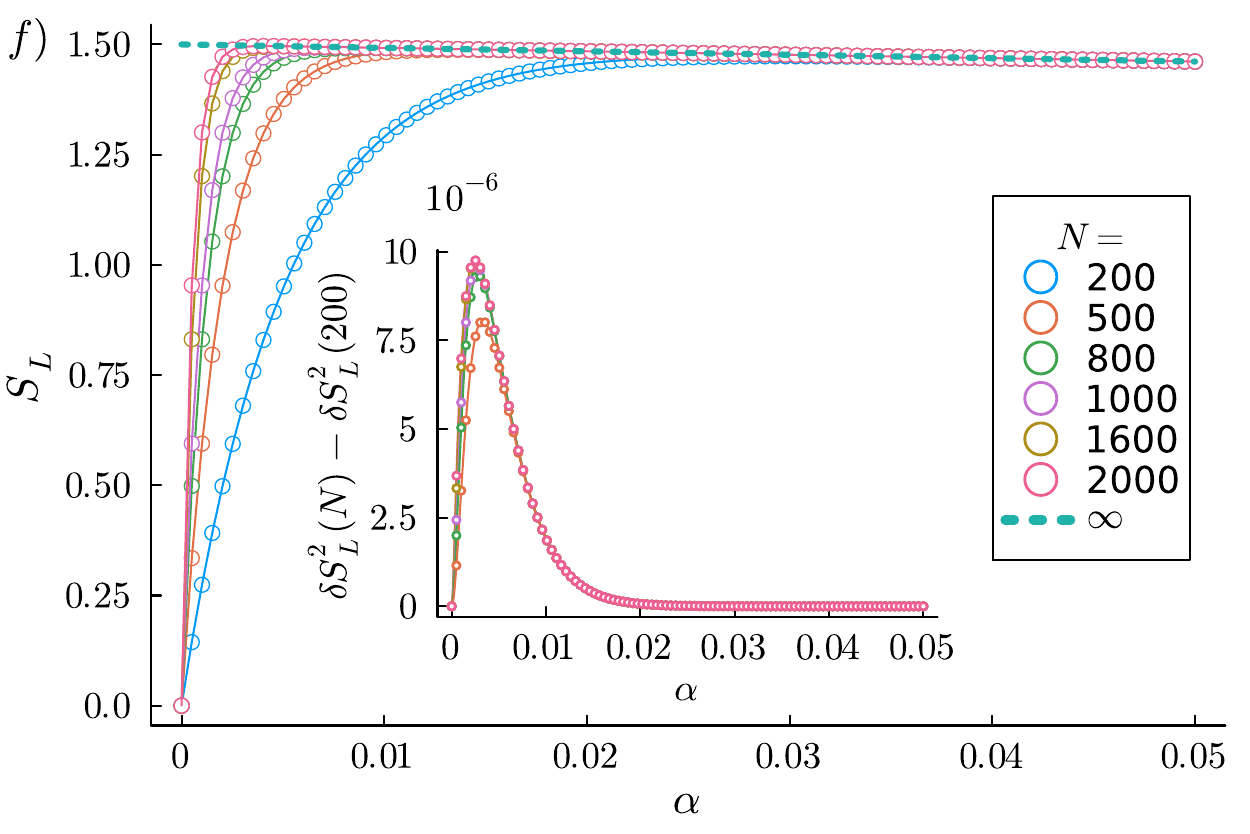}
    \label{fig:fig6}
  \end{subfigure}

  \caption{The fractionalized edge spin of magnitude $\frac{S}{2}$ for a) $S=\frac{1}{2}$, b) $S=1$, c) $S=\frac{3}{2}$, d) $S=2$, e) $S=\frac{5}{2}$ and f) $S=3$ and $\Delta=2$. The inset shows that Ansatz for variance Eq.~\eqref{ansatzvar} fits the data very well. Thus, the variance vanishes for all of these cases in the thermodynamic limit thereby showing that edge spin is a well-defined quantum observable. Note that at the right edge there is $-\frac{S}{2}$ edge spin accumulation (not shown in the figures).}
  \label{fig:all}
\end{figure}

For completeness, we have plotted the fractional edge spin accumulation $\frac{S}{2}$ for $XXZ-S$ model with $\Delta=2$ for $S=\left\{\frac{1}{2},1,\frac{3}{2},2,\frac{5}{2},3\right\}$ in Fig.~\ref{fig:all}. The inset shows the vanishing variance for each of these cases.

\begin{figure}[H]
    \centering
    \begin{subfigure}[b]{0.23\textwidth}
        \includegraphics[width=\textwidth]{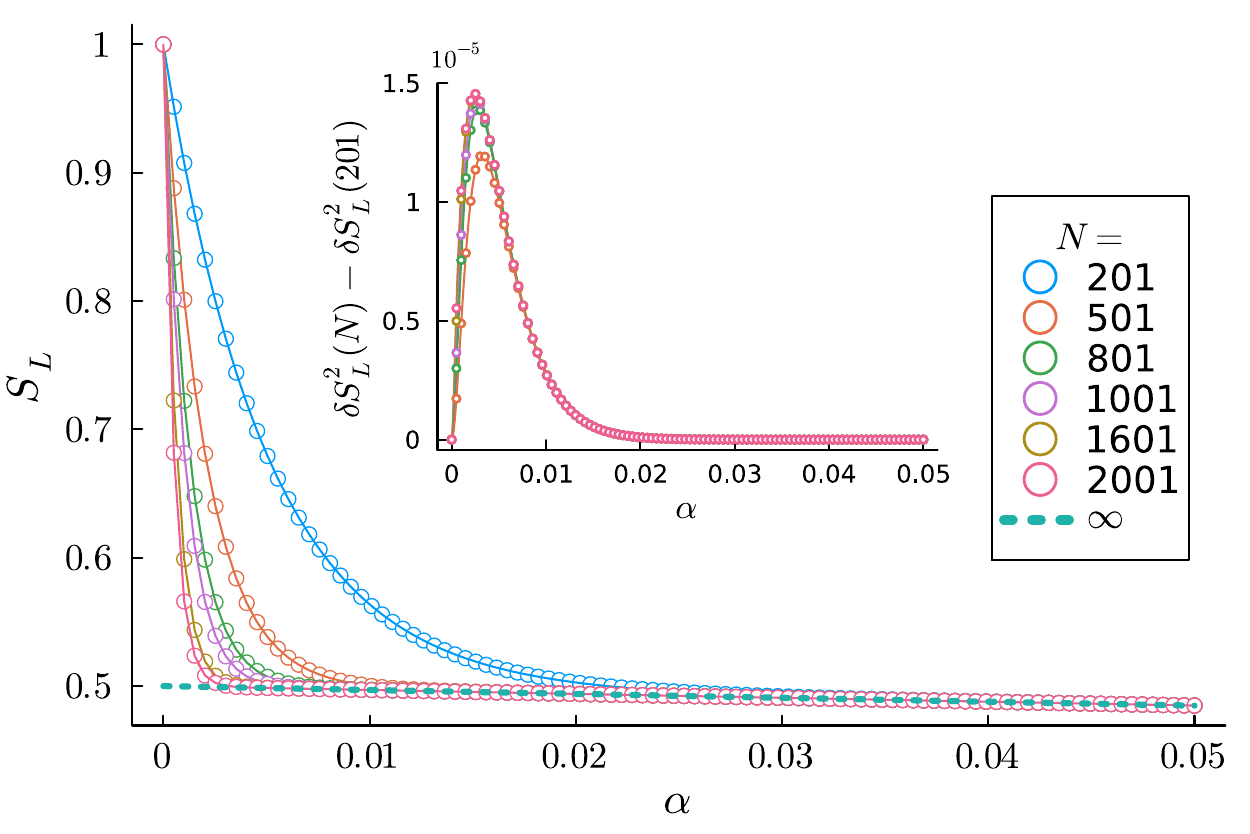}
    \end{subfigure}
    \begin{subfigure}[b]{0.23\textwidth}
        \includegraphics[width=\textwidth]{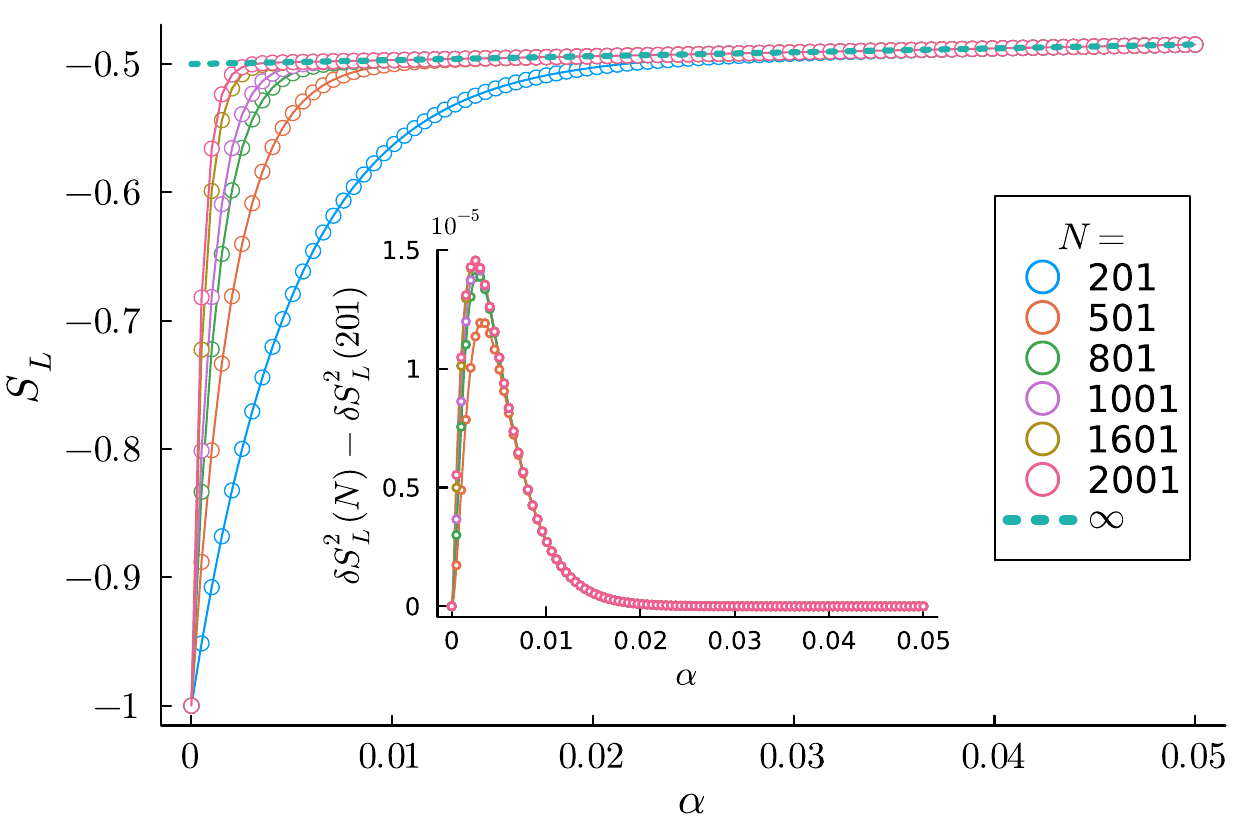}
    \end{subfigure}
    \begin{subfigure}[b]{0.23\textwidth}
        \includegraphics[width=\textwidth]{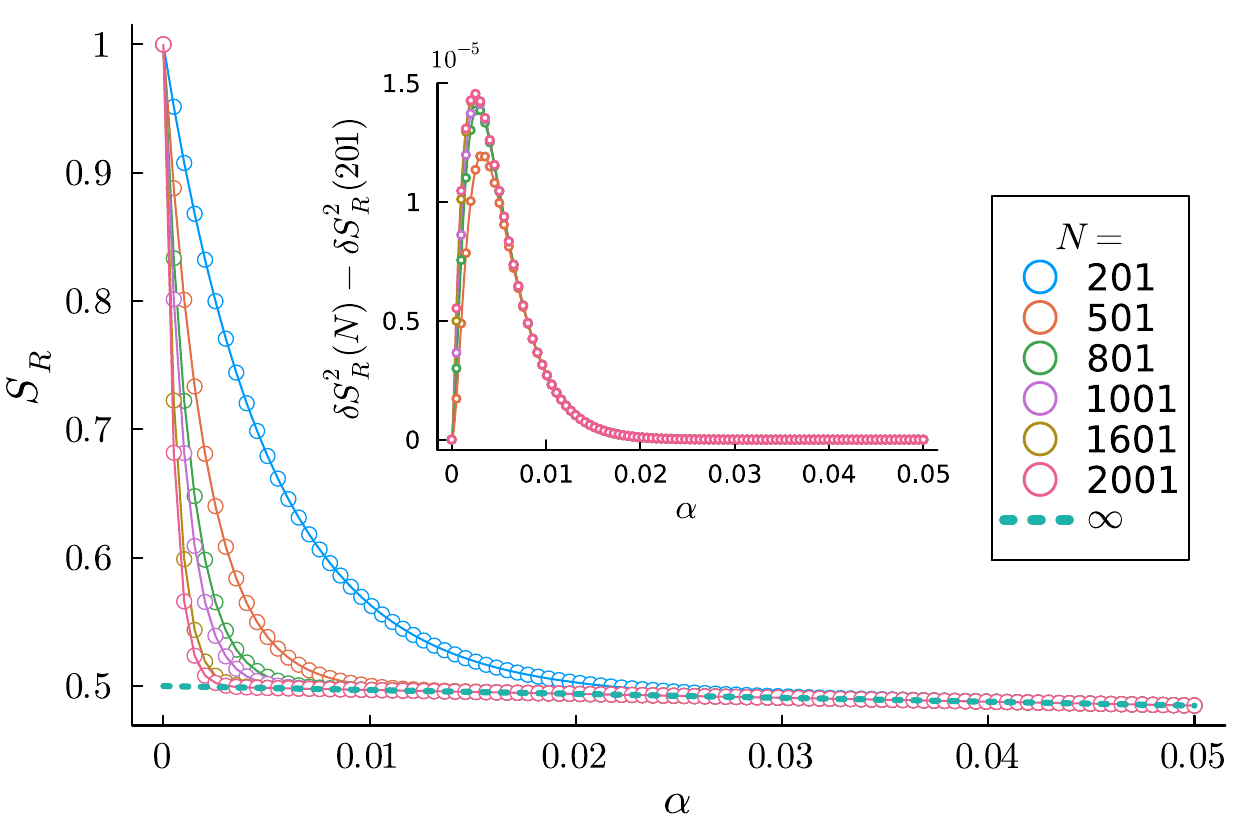}
    \end{subfigure}
    \begin{subfigure}[b]{0.23\textwidth}
        \includegraphics[width=\textwidth]{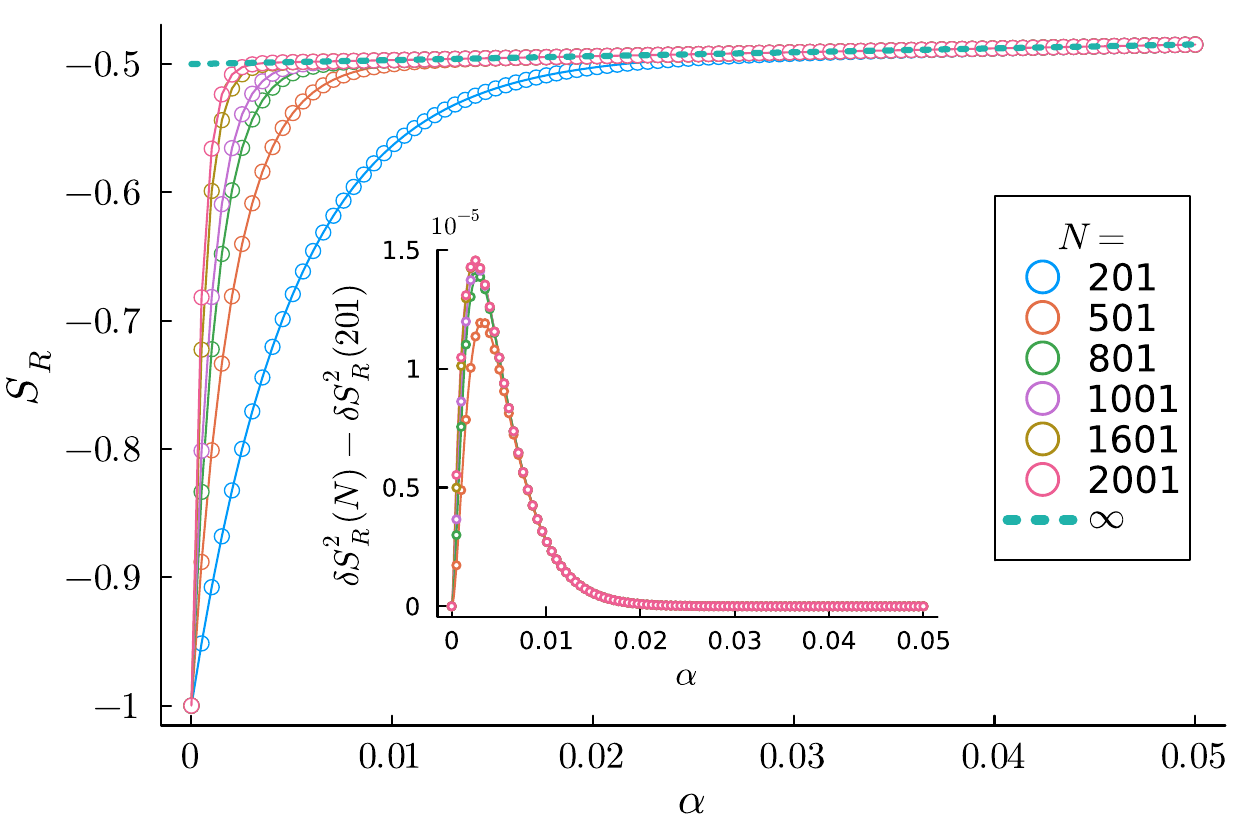}
    \end{subfigure}
    \caption{Left and right spin accumulation in the two degenerate ground state for $XXZ-1$ chain with $\Delta=2$ and odd number of sites. The inset shows that the variance fits well with the ansatz introduced in Eq.~\eqref{ansatzvar}.}
    \label{fig:all_images}
\end{figure}

Notice that the $XXZ-S$ model for $\Delta>\Delta_{c_2}$ has spontaneous symmetry breaking which implies that there are two degenerate ground states. As claimed in the main text, for the odd total number of sites, the edge spin is of the form $\left|\frac{S}{2},\frac{S}{2}\right\rangle$ and $\left|-\frac{S}{2},-\frac{S}{2}\right\rangle$. We show the representative case for $S=1$ and $\Delta=2$ in Fig.~\ref{fig:all_images}.

\begin{figure}[H]
    \centering
    \begin{subfigure}[b]{0.23\textwidth}
        \includegraphics[width=\textwidth]{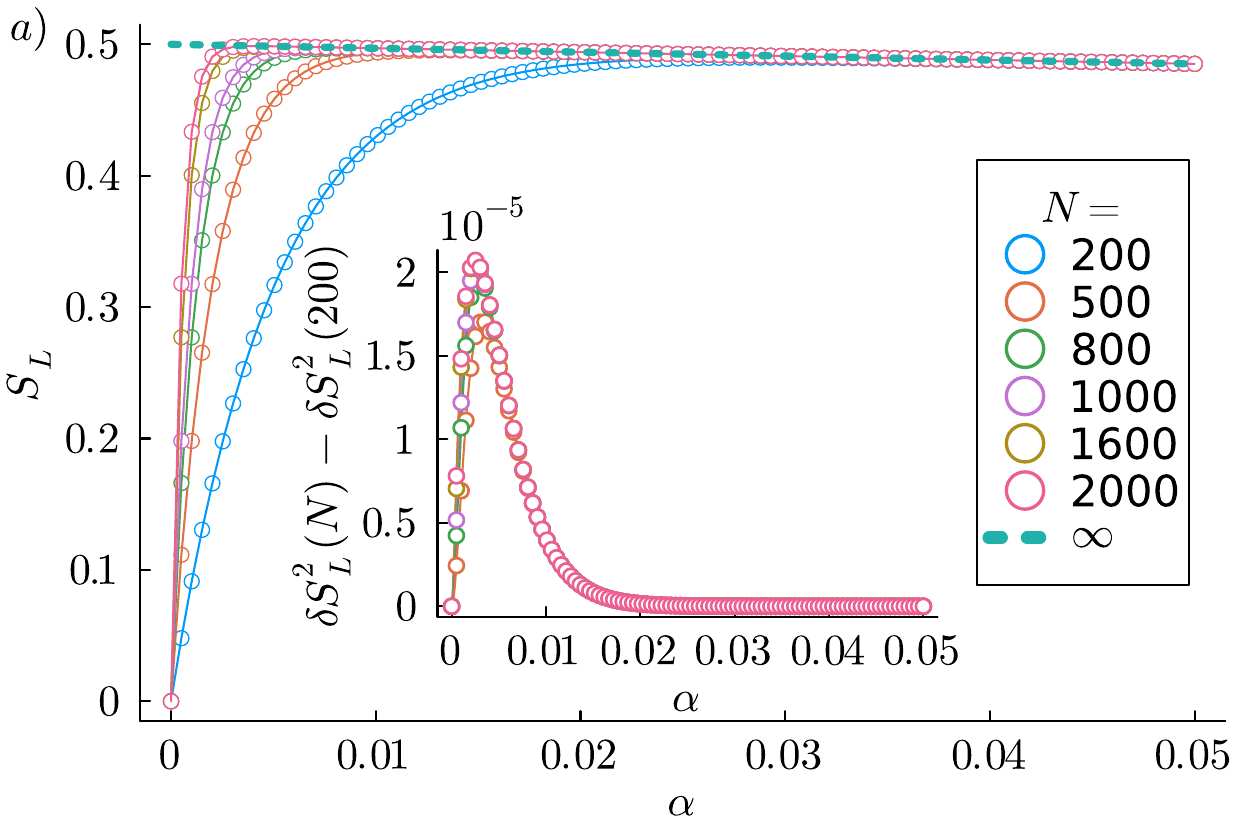}
    \end{subfigure}
    \begin{subfigure}[b]{0.23\textwidth}
        \includegraphics[width=\textwidth]{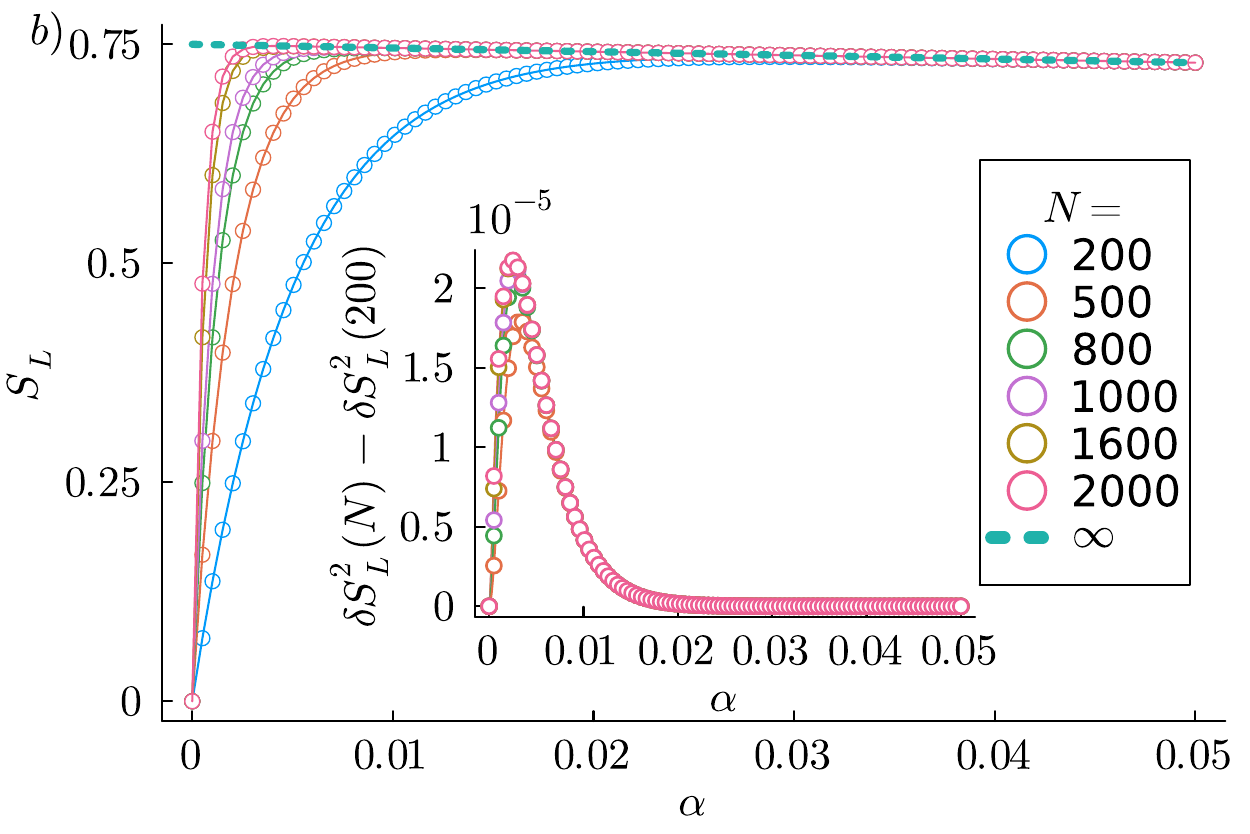}
    \end{subfigure}
    \begin{subfigure}[b]{0.23\textwidth}
        \includegraphics[width=\textwidth]{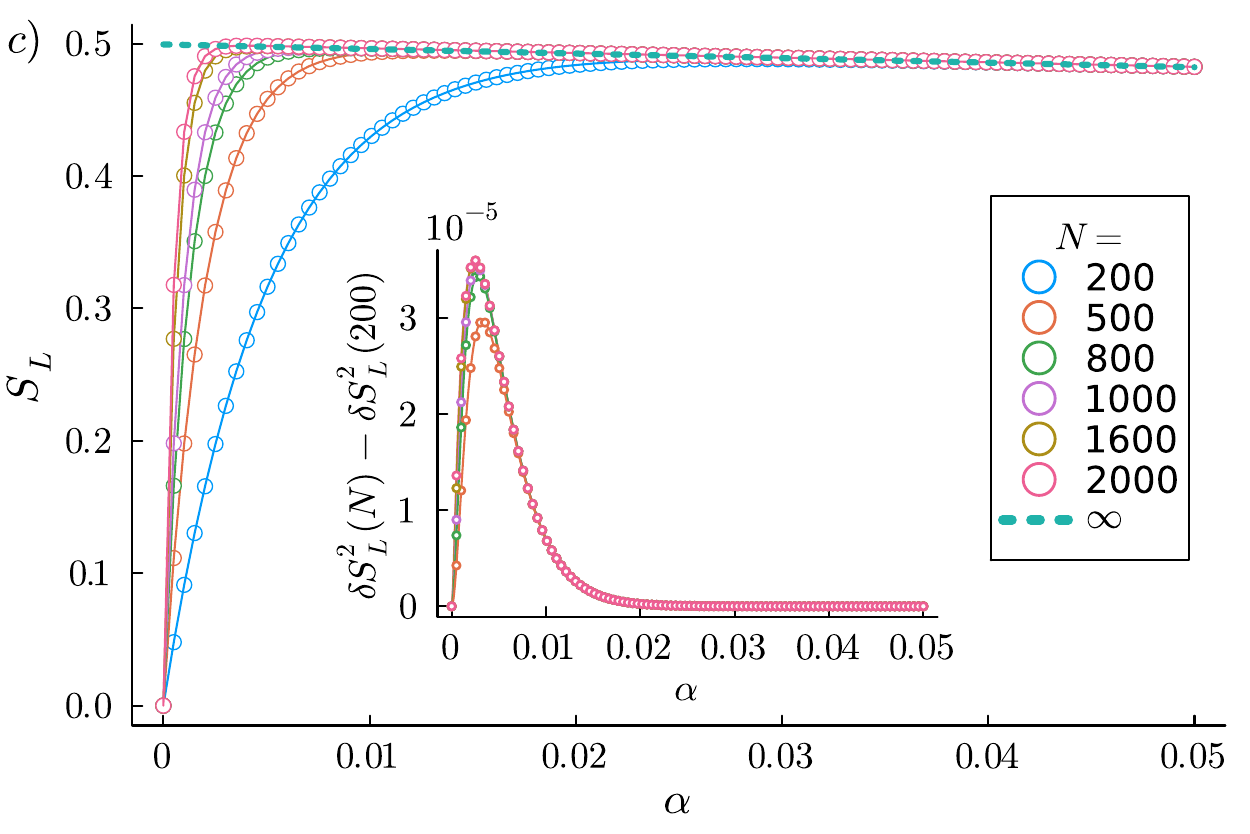}
    \end{subfigure}
    \begin{subfigure}[b]{0.23\textwidth}
        \includegraphics[width=\textwidth]{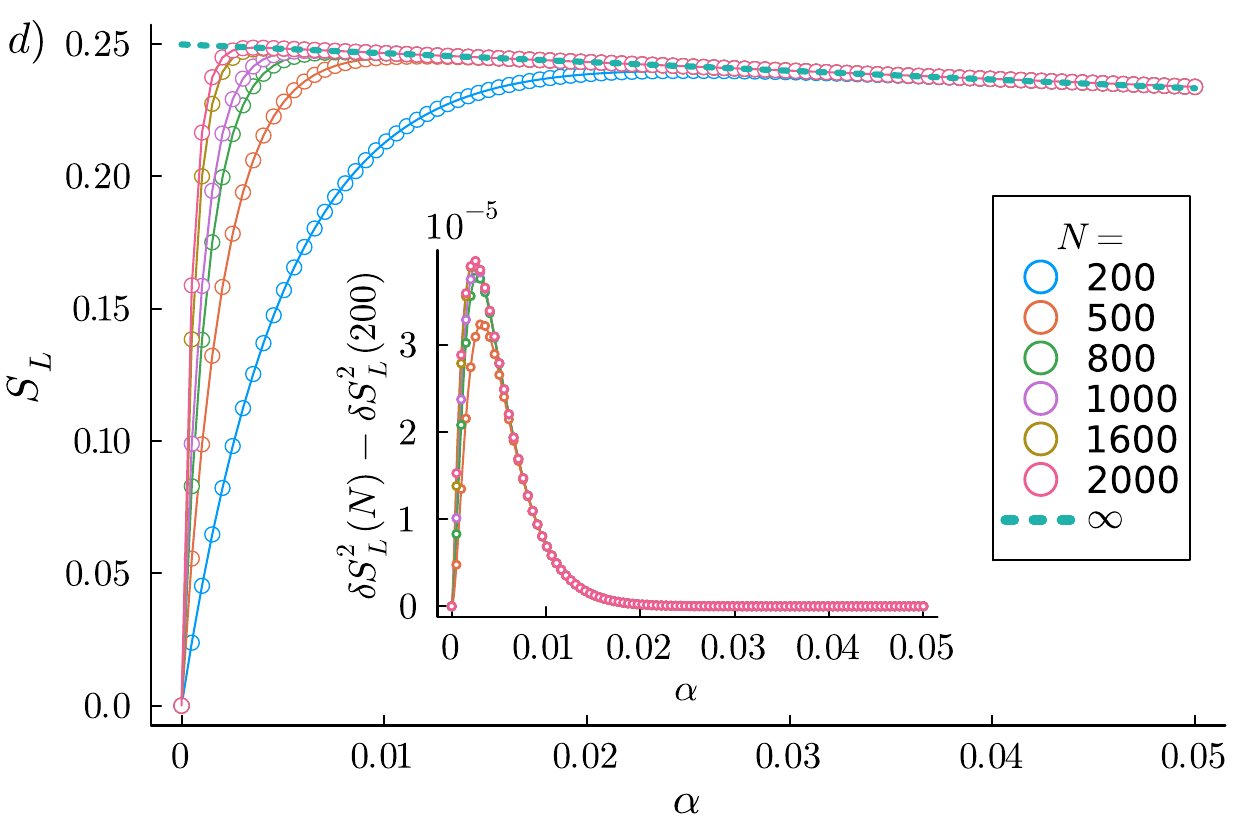}
    \end{subfigure}
    \caption{a) The fractional spin accumulation $\frac{1}{2}$ at the left end of the spin-1 anisotropic bilinear-biquadratic model for model Eq.~\eqref{biquadraticham} with
    parameter $\theta=\arctan\left(\frac{1}{3}\right)$ and $\Delta=1.5$. The inset  shows the vanishing variance of the spin operator. b) The $XXZ-\frac{3}{2}$ chain with staggered magnetic field of magnitude $h=1.25$ at the isotropic point $\Delta=1$ given by Eq.~\eqref{Hamstagg} hosts $\frac{3}{4}$ fractional edge spin. As shown in the inset, the variance of the edge spin operator vanishes in the thermodynamic limit as it fits very well with the Ansatz Eq.~\eqref{ansatzvar}. c) The fractional spin accumulation of $\frac{1}{2}$ at the left edge of the $XXZ-1$ chain given by Eq.~\eqref{HamD} with $\Delta=0.85$ and the single ion anisotropy $D=-1$. The inset shows the vanishing variance in the thermodynamic limit. d)The fractional spin accumulation of $\frac{1}{2}$ at the left edge of the $XXZ-\frac{1}{2}$ chain with next-near neighbor interaction given by Hamiltonian Eq.~\eqref{Hamj1j2} with parameters $J_1=J_2=\Delta'=1$ and $\Delta=1.75$. The inset shows the vanishing variance in the thermodynamic limit. }
    \label{explocalization1}
\end{figure}

Moreover, we show that for some perturbation respecting the three conditions, the edge mode remains robust. As example cases of the perturbation that satisfy the three conditions, we consider the biquadratic deformation
\begin{equation}
    H_{B}=\sum_{i=1}^{N-1} \cos(\theta) (\vec S_i\cdot\vec S_{i+1})_\Delta+\sin(\theta)\left(\vec S_i\cdot\vec S_{i+1} \right)^2,
    \label{biquadraticham}
\end{equation}
 a perturbation by a uniform staggered magnetic field given by 
\begin{equation}
    H_{h}= \sum_{i=1}^{N-1} (\vec S_i\cdot\vec S_{i+1})_\Delta+\sum_{i=1}^N h (-1)^i S^z_i,
    \label{Hamstagg}
\end{equation}
a perturbation by a single-ion anisotropy term
\begin{equation}
    H_D= \sum_{i=1}^{N-1} (\vec S_i\cdot\vec S_{i+1})_\Delta+D\sum_{i=1}^N (S^z_i)^2,
    \label{HamD}
\end{equation}
and a perturbation by next-near neighbor interaction
\begin{equation}
    H_{J_1J_2}= J_1\sum_{i=1}^{N-1} (\vec S_i\cdot\vec S_{i+1})_\Delta-J_2\sum_{i=1}^{N-2} (\vec S_i\cdot\vec S_{i+2})_\Delta',
    \label{Hamj1j2}
\end{equation}
in the antiferromagnetic regimes and show that the edge modes exist and the variance of the edge spin operator vanishes for representative cases in Fig.~\ref{explocalization1}.

\section{Stability of the edge modes against weak disorder}\label{sec:stability-disorder}
As mentioned in the main text, the edge modes found in this work are stable against the weak disorder satisfying our three conditions. 

\begin{figure}[H]
    \centering
    \begin{subfigure}[b]{0.23\textwidth}
        \includegraphics[width=\textwidth]{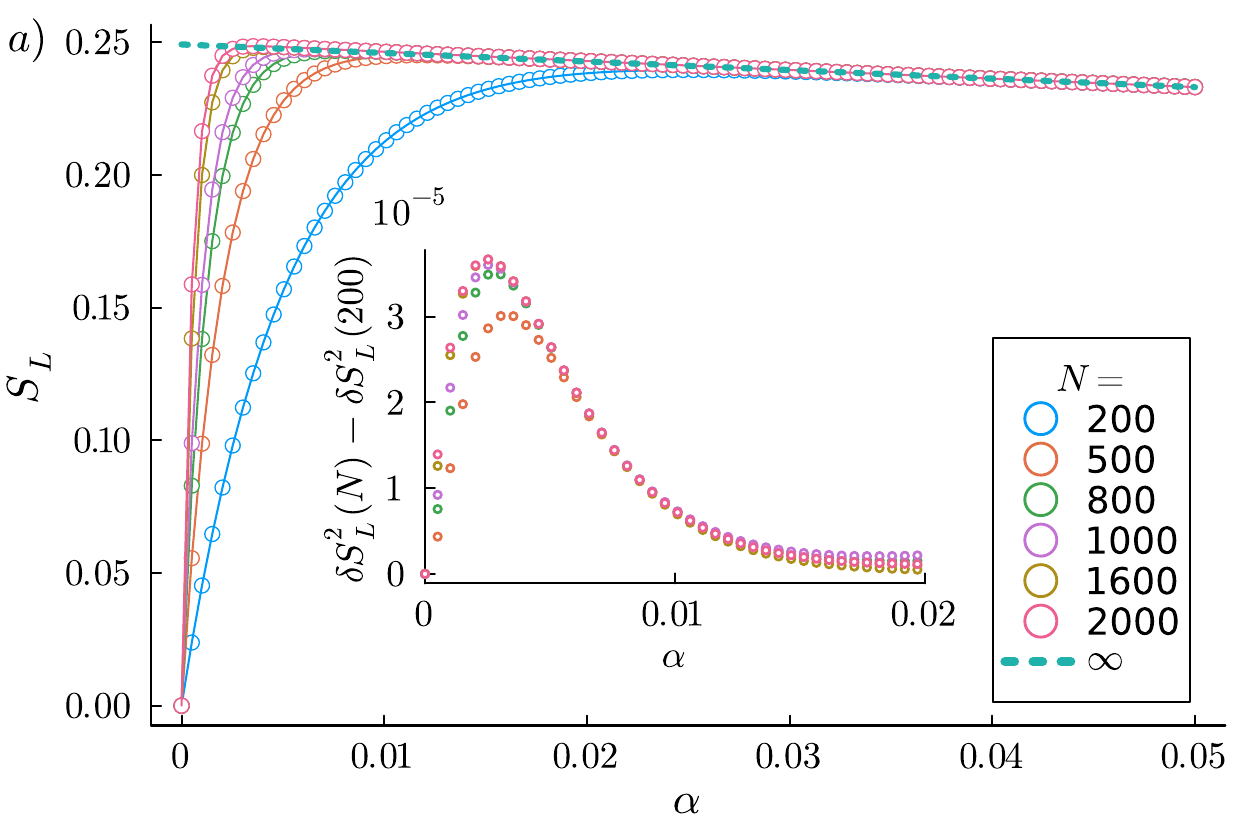}
        \label{fig:sub1}
    \end{subfigure}
    \begin{subfigure}[b]{0.23\textwidth}
        \includegraphics[width=\textwidth]{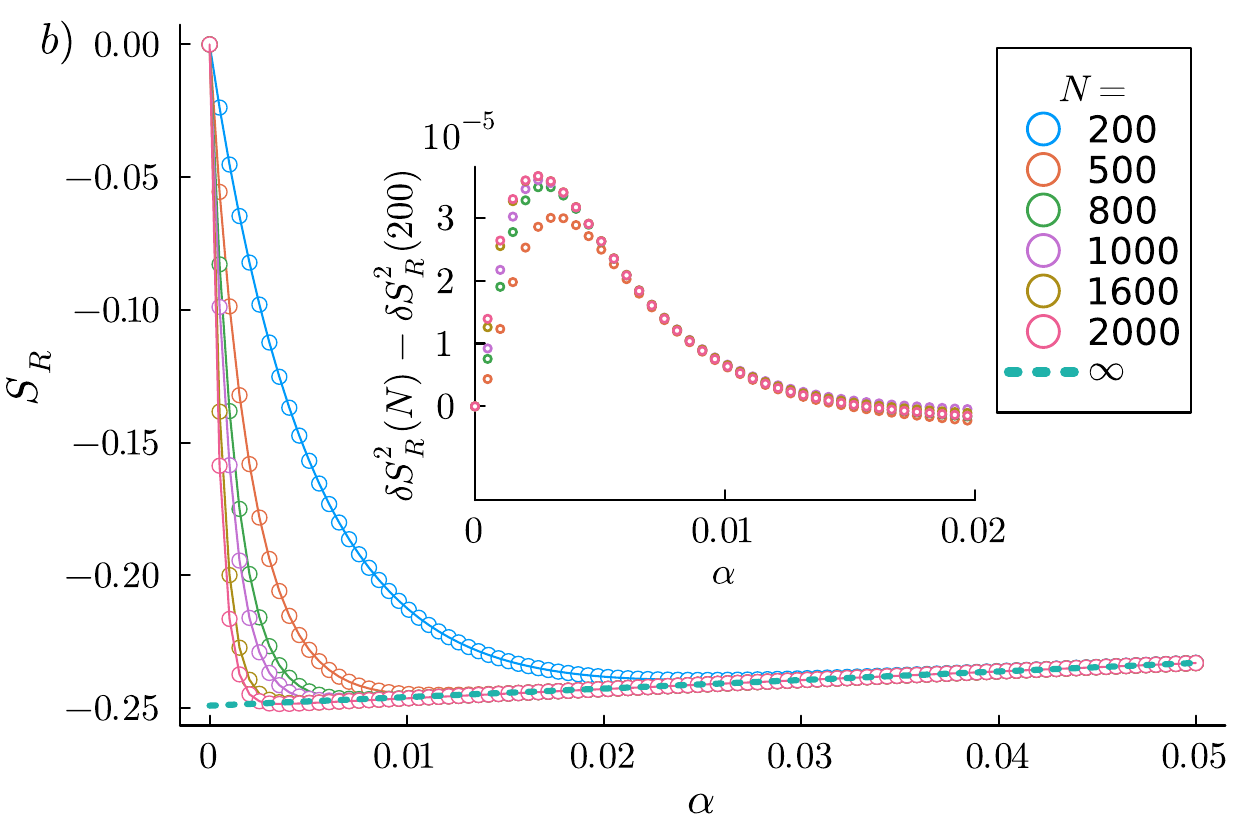}
        \label{fig:sub2}
    \end{subfigure}
    \begin{subfigure}[b]{0.23\textwidth}
        \includegraphics[width=\textwidth]{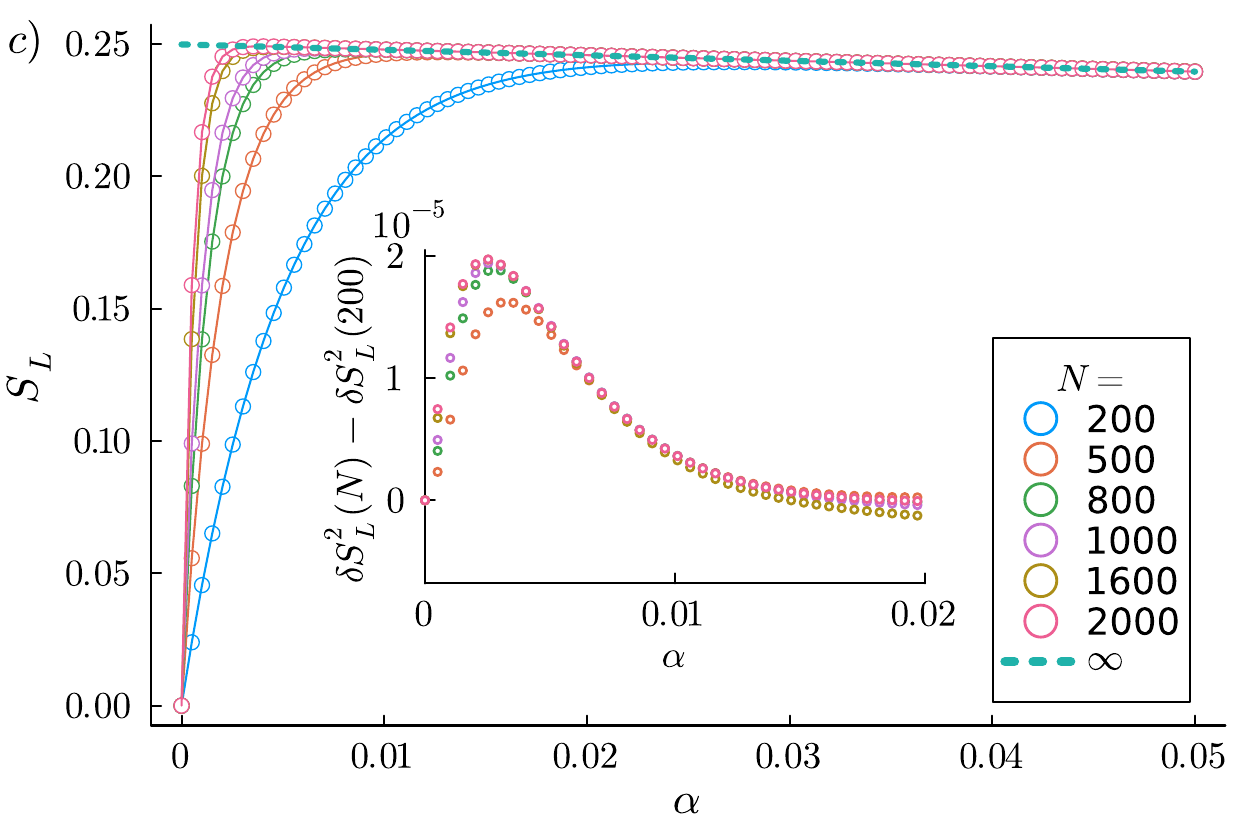}
        \label{fig:sub3}
    \end{subfigure}
    \begin{subfigure}[b]{0.23\textwidth}
        \includegraphics[width=\textwidth]{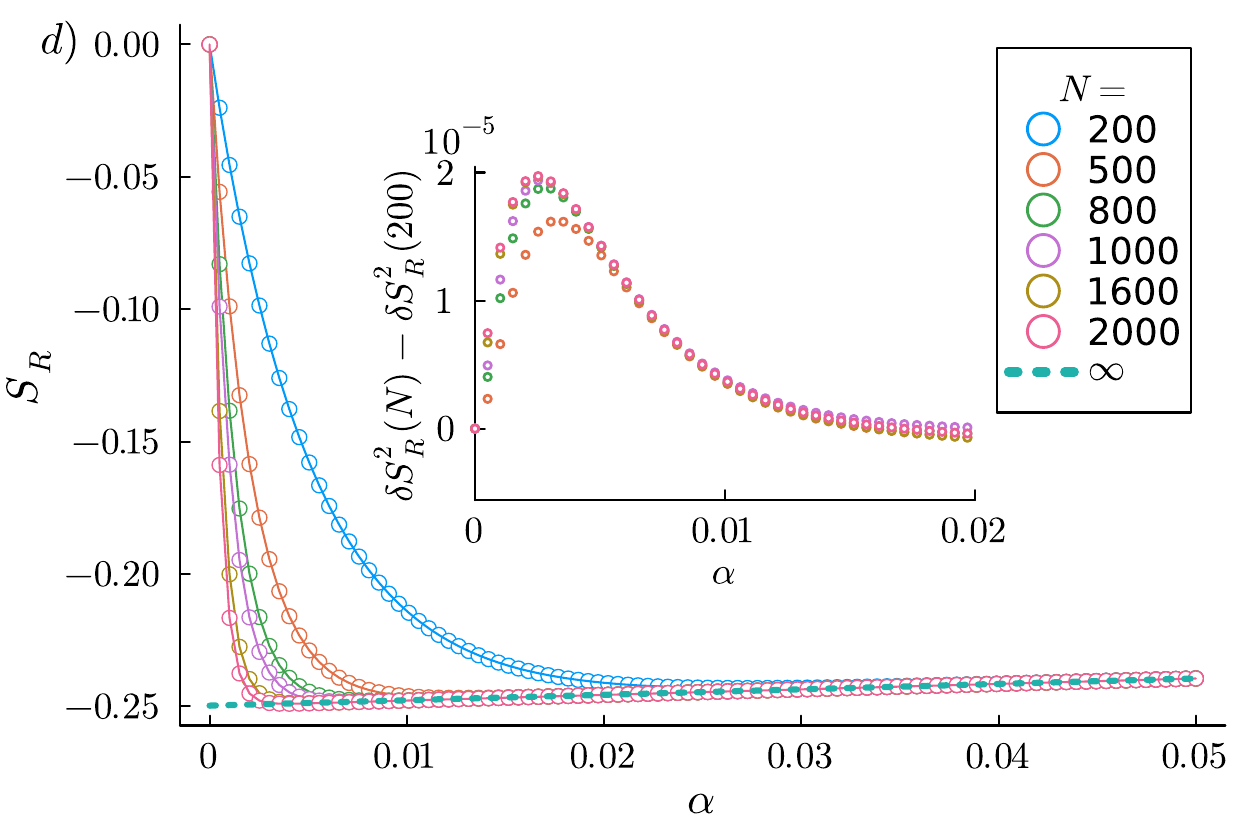}
        \label{fig:sub4}
    \end{subfigure}
    \caption{a) The left localized edge modes in the presence of weak disorder where each $W_j$ is randomly chosen from $[0,0.1]$ and $\Delta=2$. The data shown is averaged over 1000 samples. The inset shows that the variance vanishes in the thermodynamic limit. b) The right localized edge modes in the presence of weak disorder where each $W_j$ is randomly chosen from $[0,0.1]$ and $\Delta=2$. The inset shows that the variance vanishes in the thermodynamic limit. c) The left localized edge modes in the presence of disorder where each $W_j$ is randomly chosen from $[0,0.5]$ and $\Delta=2$. The inset shows that the variance vanishes in the thermodynamic limit. d) The right localized edge modes in the presence of disorder where each $W_j$ is randomly chosen from $[0,0.5]$ and $\Delta=2$. The inset shows that the variance vanishes in the thermodynamic limit.}
    \label{disorder-EM}
\end{figure}

In order to study the effect of disorder on the edge modes, we consider the Hamiltonian of the form
\begin{equation}
   H= \sum_{i=1}^{N-1} S^x_i S^x_{i+1}+S^y_i S^y_{i+1}+\Delta S^z_i S^z_{i+1}+\sum_j (-1)^j W_j S^z_j
\end{equation}
where $W_j$ are site dependent random positive magnetic fields. As shown in Fig.~\ref{disorder-EM} for the representative case of $S=\frac{1}{2}$ and $\Delta=2$, the edge modes are stable for weak disorder.

\section{Free fermion case}\label{sec:freefermion}
Here, we construct a simple analytically tractable model with spin fractionalization which can be mapped to free Fermion via Jordon-Wigner transformation. The Hamiltonian under consideration is of the form:
\begin{equation}    
H_h=\sum_{i=1}^{N-1} S^x_i S^x_{i+1} + S^y_i S^y_{i+1} + \sum_{i=1}^N h(-1)^i S^z
\end{equation}
where $S^i$ are the spin matrices for the spin-$\frac{1}{2}$. Using Jordon-Wigner transformation, the Hamiltonian can be written as
\begin{equation}
    H_h= \frac{1}{2} \left(\sum_{i=1}^{N-1} c_{i}^\dagger c_{i+1}+c_{i+1}^\dagger c_i \right)+\sum_{i=1}^N (-1)^i h (c_i^\dagger c_i-1/2)
\end{equation}

When $h=0$ and $N$ is even, the normalized wafefunction is of the form
\begin{equation}
    f_n(j)=\sqrt{\frac{2}{N+1}}\sum_j \sin\left(\frac{n\pi}{N+1} j\right)c_j^\dagger \ket{0}
\end{equation}
and the eigenvalues are
\begin{equation}
    E_n=\cos\left(k_n\right)
\end{equation}
where $k_n=\frac{n \pi}{N+1}$ and $n=\{1,\cdots, N\}$

It is important to note that for $n=j$ and $n=N+1-j$, the energy eigenvalues are negative of each other and the amplitudes of the wavefunction are the same for these pairs.

The manybody ground state, thus, can be obtained by summing all the negative modes which corresponds to
\begin{equation}
    E_{gs}=\sum_{n=\frac{N}{2}+1}^N \cos\left(\frac{n\pi}{N+1}\right)
\end{equation}

Now, when $h\neq 0$, the eigenvalues are
\begin{equation}
    E_n =\pm \sqrt{\cos^2(k)+h^2}
\end{equation}
where $k=\frac{n' \pi}{N+1}$ and $n'=\{1,\cdots,{N}/{2}\}$

and the ground state is simply
\begin{equation}
    E_{gs}(h)=-\sum_{n=1}^{\frac{N}{2}} \sqrt{h^2+\cos^2\left(\frac{n\pi}{N+1}\right)}
\end{equation}

Now, we focus on the positive parity solutions, with eigen state
\begin{equation}
    E_n=\sqrt{h^2+\cos^2\left(\frac{n\pi}{N+1}\right)}
\end{equation}
whose corresponding normalized wavefunction, is 
\begin{equation}
   \psi_n(j)=\frac{1}{\sqrt{1+\left( \frac{h}{\cos(\frac{n\pi}{N+1})+\sqrt{h^2+\cos^2\left(\frac{n\pi}{N+1}\right)}}\right)^2}}\left(f_n(j)+\frac{h}{\cos(\frac{n\pi}{N+1})+\sqrt{h^2+\cos^2\left(\frac{n\pi}{N+1}\right)}}f_{N+1-n}(j)\right)
\end{equation}
Notice, that $n$ runs from 1 to $N/2$ only. Thus, there are only half of the eigenvalues which corresponds to the positive parity solution.

Now, we consider the negative parity eigenvalues
\begin{equation}
    E_n=-\sqrt{h^2+\cos^2\left(\frac{n\pi}{N+1}\right)}
\end{equation}
Then, the eigenvalues are given by  \cite{dyachenko2021spectrum})
\begin{equation}
    \chi_n(j)=\frac{1}{\sqrt{1+\left( \frac{\cos(\frac{n\pi}{N+1})+\sqrt{h^2+\cos^2\left(\frac{n\pi}{N+1}\right)}}{h}\right)^{2}}}\left(f_n(j)-\frac{\cos(\frac{n\pi}{N+1})+\sqrt{h^2+\cos^2\left(\frac{n\pi}{N+1}\right)}}{h}f_{N+1-n}(j)\right)
\end{equation}
Again $n$ runs from 1 to $\frac{N}{2}$. Thus, there are only half of the solutions but there are precisely the one particle solutions that have negative energies! Thus, we have to use the $\chi$ solutions to do the ground state calculations!

The spin profile is simply given by
\begin{equation}
    S^z_j=\left(\sum_{n=1}^{\frac{N}{2}}|\chi_n(j)|^2\right)-1/2
    \label{spnprofile}
\end{equation}

We can simplify Eq.~\eqref{spnprofile} as
\begin{align}
    S^z_j
    &=(-1)^j\sum_{n=1}^{\frac{N}{2}}\frac{2 \sqrt{2} h \sin ^2\left(\frac{\pi  j n}{N+1}\right)}{(N+1) \sqrt{2 h^2+\cos \left(\frac{2 \pi  n}{N+1}\right)+1}}\nonumber\\
    &\stackrel{\lim_{N\to\infty}}{=}
(-1)^j\int_0^{\frac{1}{2}} \frac{2 h \sin ^2(\pi  j x)}{\sqrt{h^2+\cos ^2(\pi  x)}} \, dx
    \label{intrep}
\end{align}
Note that in the last step, we took $N\to\infty$ limit. Fortunately, this does not remove the effect of the boundary in the left end. But indeed, it washes away the effect on the right boundary. However, if we recall $S^z_j=-S^z_{N+1-j}$ for even parity and $S^z_j=S^z_{N+1-1}$ for odd parity, we shall be able to construct the full profile of the chain.

It is now possible to find the bulk antiferromagnetic order as
\begin{equation}
    \mathfrak{S}(h)=\int_0^{\frac12} \frac{h}{\sqrt{h^2+\cos^2(\pi x)}}\mathrm{d}x=\frac{h K\left(\frac{1}{h^2+1}\right)}{\pi  \sqrt{h^2+1}}
\end{equation}
where $K$ is the complete elliptic integral of the first kind defined as
\begin{equation}
   K(k) = \int_0^{\frac{\pi}{2}} \frac{d\theta}{\sqrt{1-k \cos^2\theta}}
\end{equation}

Although, a closed form expression for Eq.~\eqref{intrep} for generic $j$ seems difficult, it is not difficult to numerically observe that
\begin{equation}
    \sum_{j=1}^{\frac{N}{2}} S^z_j -\frac{1}{2} \mathfrak{S}(h)=-\frac{1}{4}
\end{equation}
irrespective of the value of $h$ for even $N$.

\section{Integrable spin-S chain}\label{sec:spin1}

Recall that the basic building block of anisotropic spin-$\frac12$ Heisenberg spin chain is the $R$ matrix
\begin{equation}
   R^{\frac{1}{2},\frac{1}{2}}(u)=\left(
\begin{array}{cccc}
 \frac{\sinh (\eta +u)}{\sinh(\eta )} & 0 & 0 & 0 \\
 0 & \frac{\sinh (u)}{\sinh(\eta )} & 1 & 0 \\
 0 & 1 & \frac{\sinh (u)}{\sinh(\eta )} & 0 \\
 0 & 0 & 0 & \frac{\sinh (\eta +u)}{\sinh(\eta )} \\
\end{array}
\right) 
\label{rmat}
\end{equation}

    which is  a solution of the Yang-Baxter equation
\begin{equation}
    R_{12}(u-v)R_{13}(u)R_{23}(v)=R_{23}(v)R_{13}(u)R_{12}(u-v).
\end{equation}
Here the superscript in $R$ in Eq.~\eqref{rmat} denotes that both the auxiliary space and the physical space are spin-$\frac{1}{2}$ (or in other words, they are two dimensional).

Now, we construct the $R^{j,s}(u)$, where the auxiliary space is $2j+1$ dimensional and the physical space is $2s+1$ dimensional by using the symmetric fusion method described in \cite{wang2015off}, 
such that the fused $R$ matrix can be written as
\begin{equation}
    R^{j,s}(u)  = P_{\{1 \cdots 2 j\}}^{+} \prod_{k=1}^{2 j}\left\{R^{\frac{1}{2},s}\left(u+\left(k-j-\frac{1}{2}\right) \eta\right)\right\} P_{\{1 \cdots 2 j\}}^{+}
\end{equation}
where
    \begin{equation}
    R^{\frac{1}{2}, s} (u) = P_{\{1 \cdots 2 s\}}^{+} \prod_{k=1}^{2 s}\left\{R^{\frac12,\frac12}\left(u+\left(k-\frac{1}{2}-s\right) \eta\right)\right\} P_{\{1 \cdots 2 s\}}^{+}\;,
\end{equation}

and $ P_{\{1 \cdots 2 s\}}^{+}$ is a symmetric projector 
\begin{equation}
     P_{\{1 \cdots 2 s\}}^{+}= \frac{1}{(2 s) !} \prod_{k=1}^{2 s}\left(\sum_{l=1}^k \mathbf{P}_{l k}\right)\;.
\end{equation}
with $\mathbf{P}_{l k}$ being the permutation operator.

Let us consider the following two single row transfer matrices
 \begin{align}
T_0^{j,s}(u) &= R_{0,N}^{j,s}(u-\theta_{N})R_{0,N-1}^{j,s}(u-\theta_{N-1})\cdots R_{0,2}^{j,s}(u-\theta_2)R_{0,1}^{j,s}(u-\theta_1)\nonumber\\
\hat{T}_0^{j,s}(u) &= R_{0,1}^{j,s}(u+\theta_1)R_{0,2}^{j,s}(u+\theta_2)\cdots R_{0,N-1}^{j,s}(u+\theta_{N-1})R_{0,N}^{j,s}(u+\theta_{N})\nonumber
\end{align}

Now, we define the monodromy matrix
\begin{equation}
    \Xi^{j,s}(u)=T_0^{j,s}(u)\hat T^{j,s}(u)
\end{equation}
The trace of monodromy matrix over the auxiliary space is defined as the double row transfer matrix
\begin{equation}
    t(u)=\operatorname{tr}_0\Xi^{j,s}(u)
    \label{tmat}
\end{equation}
    
Using the fusion hierarchy of the transfer matrices, one can show that the eigenvalues $\Lambda(u)$ of the transfer matrix matrix $t(u)$ satisfy Baxter’s $T-Q$ relation \cite{baxter1972partition,baxter2000partition,wang2015off} 
such that regularity condition on the $T-Q$ relation gives the Bethe equation. Following the standard procedure described in detail in \cite{wang2015off}, we obtain the BAE for integrable spin-S chain as
\begin{equation}
     \begin{gathered}
\left(\frac{\sin \frac{1}{2}\left(\lambda _j-{2i \eta S}\right)}{\sin \frac12\left(\lambda _j+{2i \eta S}\right)}\right)^{2 N}
\frac{\cos^2\frac12\left(\lambda _j+{2i \eta S}\right)}{\cos^2\frac12\left(\lambda _j-{2i \eta S}\right)} =\prod_{k=1(\neq j)}^M \frac{\sin\frac12 \left(\lambda_j-\lambda_k-2i\eta \right) \sin \frac12\left(\lambda_j+\lambda_k-2i\eta \right)}{\sin\frac12 \left(\lambda_j-\lambda_k+2i\eta \right) \sin \frac12\left(\lambda_j+\lambda_k+2i\eta \right)}
\end{gathered}
\label{spin1eqn}
\end{equation}

The integrable Hamiltonian is now obtained by taking log derivative of the transfer matrix Eq.~\eqref{tmat}. 

For $S=\frac{1}{2}$, one obtains a clean $XXZ$ chain of the form
\begin{equation}
    H_{\frac{1}{2}}=\sum_{i=1}^{N-1}\frac{J}{2}(\sigma_i^x\otimes \sigma_{i+1}^x+\sigma_i^y\otimes \sigma_{i+1}^y+\Delta \sigma_i^z\otimes \sigma_{i+1}^z),
\end{equation}

but when $S>\frac{1}{2}$, there are higher order spin interactions. Here, we write the Hamiltonian for spin-1 case explicitly \cite{frappat2007complete}

\begin{equation}
H_1=\sum_n \sigma_{n} - (\sigma_{n})^{2} + 2\sinh^{2}(\eta) \left[ \sigma^{z}_{n} + (S^{z}_{n})^{2} + (S^{z}_{n+1})^{2} - (\sigma^{z}_{n})^{2} \right] - 4\sinh^{2}\left(\frac{\eta}{2}\right) (\sigma^{\perp}_{n} \sigma^{z}_{n} + \sigma^{z}_{n} \sigma^{\perp}_{n}),
\end{equation}

where
\begin{align*}
\sigma_{n} &= \vec{S}_{n} \cdot \vec{S}_{n+1} \\
\sigma^{\perp}_{n} &= S^{x}_{n} S^{x}_{n+1} + S^{y}_{n} S^{y}_{n+1} \\
\sigma^{z}_{n} &= S^{z}_{n} S^{z}_{n+1}
\end{align*}
and $S_n^i$ for $i=\{x,y,z\}$ are the spin-1 representation of $SU(2)$.

We can write the model as
\begin{equation}
    H_1=H_\Delta + H^I
\end{equation}
where 
\begin{equation}
    H_\Delta=\sum_n S^x_n S^x_{n+1}+S^y_n S^y_{n+1}+\Delta S^z_n S^z_{n+1},
\end{equation}
with $\Delta=1+2\sinh^2\eta$ and
the integrable deformation terms
\begin{equation} \label{fused}
    H^I=\sum_n  2\sinh^{2}(\eta) \left[ (S^{z}_{n})^{2} + (S^{z}_{n+1})^{2} - (\sigma^{z}_{n})^{2} \right] - 4\sinh^{2}\left(\frac{\eta}{2}\right) (\sigma^{\perp}_{n} \sigma^{z}_{n} + \sigma^{z}_{n} \sigma^{\perp}_{n})-(\sigma_n)^2.
\end{equation}

Notice, that the ground state of spin-S chain is given by a vacuum of 2S string solution. This means that for $S=1/2$, the ground state is made up of $1-$string solution whereas for $S=1$, it is made up of $2-$string solution. The detail solution of $S=1/2$ can be found in \cite{pasnoori2023spin} and here we show it for an example case of $S=1$ where by substituting $2-$string, $\lambda_j\to \chi_j\pm i\eta$, we obtain

\begin{align}
&\left(\frac{\sin \left(\frac{1}{2} \left(\chi _j-3 i \eta \right)\right)}{\sin \left(\frac{1}{2} \left(\chi _j+3 i \eta \right)\right)}\right)^{2 N} \left(\frac{\sin \left(\frac{1}{2} \left(\chi _j-i \eta \right)\right)}{\sin \left(\frac{1}{2} \left(\chi _j+i \eta \right)\right)}\right)^{2 N}\left(\frac{\cos \left(\frac{1}{2} \left(\chi _j+3 i \eta \right)\right)}{\cos \left(\frac{1}{2} \left(\chi _j-3 i \eta \right)\right)}\right)^2 \left(\frac{\cos \left(\frac{1}{2} \left(\chi _j+i \eta \right)\right)}{\cos \left(\frac{1}{2} \left(\chi _j-i \eta \right)\right)}\right)^2\nonumber\\
&\frac{\sin \left(\chi _j- i \eta \right)}{\sin \left(\chi _j+ i \eta \right)}=\prod_{k=1}^M \frac{\sin \left(\frac{1}{2} \left(\chi _j-\chi _k-4 i \eta\right)\right)}{\sin \left(\frac{1}{2} \left(\chi _j-\chi _k+4 i \eta \right)\right)}\left(\frac{\sinh \left(\frac{1}{2} \left(\chi _j-\chi _k-2 i \eta \right)\right)}{\sinh \left(\frac{1}{2} \left(\chi _j-\chi _k+2 i \eta\right)\right)}\right)^2\times \nonumber\\
&\hspace{5cm}\frac{\sin \left(\frac{1}{2} \left(\chi _j+\chi _k-4 i \eta\right)\right)}{\sin \left(\frac{1}{2} \left(\chi _j+\chi _k+4 i \eta \right)\right)}\left(\frac{\sinh \left(\frac{1}{2} \left(\chi _j+\chi _k-2 i \eta \right)\right)}{\sinh \left(\frac{1}{2} \left(\chi _j+\chi _k+2 i \eta\right)\right)}\right)^2
\end{align}

Writing,
\begin{equation}
    \frac{\sin \left(\chi _j- i \eta \right)}{\sin \left(\chi _j+ i \eta \right)}= \frac{\sin \left(\frac{1}{2} \left(\chi _j- i \eta \right)\right) \cos \left(\frac{1}{2} \left(\chi _j- i \eta \right)\right)}{\sin \left(\frac{1}{2} \left(\chi _j+ i \eta \right)\right) \cos \left(\frac{1}{2} \left(\chi _j+i \eta \right)\right)},
\end{equation}

we rewrite the Bethe equations in the convenient logarithmic form as

\begin{align}
   &(2N+1) \phi\left(\chi_j,\eta\right)+2N\phi(\chi_j,3\eta)-2\psi\left(\chi_j,\eta\right)-2\psi(\chi_j,3\eta)+\psi(\chi_j,\eta)\nonumber\\
   &=\pi i I_j +\sum_{k}\left[\phi(\chi_j+\chi_k,4\eta)+\phi(\chi_j-\chi_k,4\eta)+ 2 \phi(\chi_j-\chi_k,2\eta)+2 \phi(\chi_j+\chi_k,2\eta) \right]
   \label{logBAE}
\end{align}

Where we introduced
\begin{align}
    \phi(a,b)&=\log\left( \frac{\sin\frac12(a-ib)}{\sin\frac12(a+ib)}\right)\\
    \psi(a,b)&=\log\left( \frac{\cos\frac12(a-ib)}{\cos\frac12(a+ib)}\right)
\end{align}

We extract the density of roots in the ground state $\rho_0(\lambda)$ by subtracting Eq.~\eqref{logBAE} written for $\lambda_j$ from the same equation written for $\lambda_{j+1}$ and expanding in the difference $\Delta\lambda=\lambda_{j+1}-\lambda_j$. This gives
\begin{align}
    (2N+1) &a(\chi ,\eta )+2Na(\chi,3\eta)-2 a\left(\chi -\pi,\eta \right)-2 a\left(\chi -\pi,3\eta \right)+ a(\lambda-\pi,\eta) \nonumber\\
    &=2\pi \rho(\lambda)+\int \rho(\lambda) \left[a(\lambda-\lambda',4\eta)+a(\lambda+\lambda',4\eta)+2a(\lambda-\lambda',2\eta)+2a(\lambda+\lambda',2\eta) \right]\mathrm{d}\lambda'\nonumber\\
    &+4\pi\delta(\lambda)+4\pi\delta(\lambda-\pi),
    \label{rtdens}
\end{align}

 where we introduced
\begin{equation}
    a(x,y)=\frac{\sinh(y)}{ \cosh(y)- \cos(x)}
\end{equation}
and we added delta function at $\lambda=0$ and $\lambda=\pi$ to remove the two solutions which lead to vanishing wavefunction.

Using the following convention for Fourier transform
\begin{equation}
    f(x)=\sum_{\omega=-\infty}^{\infty} \hat{f}(\omega) e^{i \omega x} \quad\quad \text{and}\quad\quad \hat{f}(\omega)=\frac{1}{2 \pi} \int_{-\pi}^\pi f(x) e^{-i \omega x} d x,
\end{equation}
we compute
\begin{equation}
    \hat a(\omega,y)=\frac{1}{2\pi}\int_{-\pi }^\pi e^{-i\omega x}\sum _{k=-\infty }^{\infty } e^{i k x } \left(e^{- y}\right)^{| k| }=e^{-y|\omega|}
\end{equation}
Now the solution of Eq.~\eqref{rtdens} is immediate in the Fourier space
\begin{equation}
    \tilde\rho_{\ket{1}}(\omega)=\frac{(2N+1)e^{-\eta|\omega|}+2Ne^{-3\eta|\omega|}+(-1)^\omega e^{-\eta|\omega|}-2(-1)^\omega e^{-\eta|\omega|}-2(-1)^\omega e^{-3\eta|\omega|}-(1+(-1)^\omega)}{4\pi(1+e^{-4\eta|\omega|}+2e^{2\eta|\omega|})}
\end{equation}
The total number of center of roots is given by
\begin{equation}
    M_{\ket{1}}=\int_{-\pi}^\pi \rho_{\ket{\frac12}}(\lambda)\mathrm{d}\lambda=2\pi \tilde \rho_{\ket{\frac12}}(0)=\frac{N-1}{2}
\end{equation}

However, this is not possible for even $N$. If this were a valid state, the spin of this state would be
\begin{equation}
   S^z= N-2M=1
\end{equation}

Once again due to the $SU(2)$ symmetry, there is also configuration with same Bethe roots with spin $-1$.

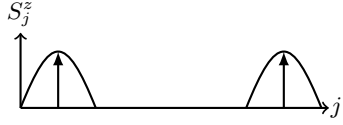
\begin{figure}[H]
    \centering
    \begin{tikzpicture}
    \draw[thick,->] (-2,0) -- (2.1,0);
    \draw[-latex, thick] (-1.5,0) -- (-1.5,0.75);
     \draw[-latex, thick] (1.5,0) -- (1.5,0.75);
    \draw[thick] (2,0) .. controls (1.6,1) and (1.4,1) .. (1,0);
     \draw[thick] (-2,0) .. controls (-1.6,1) and (-1.4,1) .. (-1,0);
     \draw[->,thick] (-2,0) --(-2,1);
     \node at (-2,1.25) {$S^z_j$};
     \node at (2.2,0) {$j$};
\end{tikzpicture}
    \caption{Schematic of the edge localized spin quarter $S/2$.}
    \label{fig:wcong}
\end{figure}

It will be important later to understand the spin configuration of this configuration of the Bethe roots. The total spin accumulation $S^z=1$ comes from the two boundary terms equally contributing to this spin accumulation. Due to the gap in the bulk, this spin configuration has to be $\frac{1}{2}$ in each end of the chain. Moreover, because of the $SU(2)$ symmetry there is also configuration with the same Bethe root distribution where the total spin is $-1$. These two doubly degenerate root configurations (which are not valid state) have total
 spin $\pm 1$ where each end has $\pm \frac{1}{2}$ sharply localized spin as shown in Fig.~\ref{fig:wcong}.

Now, to construct a valid ground state state, we need to add the boundary string. Adding the boundary string $\pi\pm 2i \eta$ the list of the solution in Eq.~\eqref{spin1eqn}, we obtain the equation for ground state as

\begin{align}
\left(\frac{\sin \left(\frac{1}{2} \left(\chi _j-3 i \eta \right)\right)}{\sin \left(\frac{1}{2} \left(\chi _j+3 i \eta \right)\right)}\right)^{2 N} &\left(\frac{\sin \left(\frac{1}{2} \left(\chi _j-i \eta \right)\right)}{\sin \left(\frac{1}{2} \left(\chi _j+i \eta \right)\right)}\right)^{2 N}\left(\frac{\cos \left(\frac{1}{2} \left(\chi _j+3 i \eta \right)\right)}{\cos \left(\frac{1}{2} \left(\chi _j-3 i \eta \right)\right)}\right)^2 \left(\frac{\cos \left(\frac{1}{2} \left(\chi _j+i \eta \right)\right)}{\cos \left(\frac{1}{2} \left(\chi _j-i \eta \right)\right)}\right)^2\nonumber\\
\frac{\sin \left(\frac{1}{2} \left(\chi _j- i \eta \right)\right) \cos \left(\frac{1}{2} \left(\chi _j- i \eta \right)\right)}{\sin \left(\frac{1}{2} \left(\chi _j+ i \eta \right)\right) \cos \left(\frac{1}{2} \left(\chi _j+i \eta \right)\right)}&= \left(\frac{\cos \left(\frac{1}{2} \left(\chi _j-5 i \eta \right)\right)}{\cos \left(\frac{1}{2} \left(\chi _j+5 i \eta \right)\right)}\right)^2 \left(\frac{\cos \left(\frac{1}{2} \left(\chi _j-3 i \eta \right)\right)}{\cos \left(\frac{1}{2} \left(\chi _j+3 i \eta \right)\right)}\right)^2\nonumber\\
&\prod_{k=1}^M \frac{\sin \left(\frac{1}{2} \left(\chi _j-\chi _k-4 i \eta\right)\right)}{\sin \left(\frac{1}{2} \left(\chi _j-\chi _k+4 i \eta \right)\right)}\left(\frac{\sinh \left(\frac{1}{2} \left(\chi _j-\chi _k-2 i \eta \right)\right)}{\sinh \left(\frac{1}{2} \left(\chi _j-\chi _k+2 i \eta\right)\right)}\right)^2\times\nonumber\\
&\frac{\sin \left(\frac{1}{2} \left(\chi _j+\chi _k-4 i \eta\right)\right)}{\sin \left(\frac{1}{2} \left(\chi _j+\chi _k+4 i \eta \right)\right)}\left(\frac{\sinh \left(\frac{1}{2} \left(\chi _j+\chi _k-2 i \eta \right)\right)}{\sinh \left(\frac{1}{2} \left(\chi _j+\chi _k+2 i \eta\right)\right)}\right)^2
\end{align}

\begin{figure}[H]
    \centering
    \centering
    \begin{tikzpicture}
    \draw[thick,->] (-2,0) -- (2.1,0);
    \draw[-latex, thick] (-1.5,0) -- (-1.5,0.75);
     \draw[-latex, thick] (1.5,0) -- (1.5,-0.75);
    \draw[thick] (2,0) .. controls (1.6,-1) and (1.4,-1) .. (1,0);
     \draw[thick] (-2,0) .. controls (-1.6,1) and (-1.4,1) .. (-1,0);
      \draw[->,thick] (-2,0) --(-2,1);
     \node at (-2,1.25) {$S^z_j$};
     \node at (2.2,0) {$j$};

        \draw[thick, ->] (4,0) -- (8.1,0);
    \draw[-latex, thick] (4.5,0) -- (4.5,-0.75);
     \draw[-latex, thick] (7.5,0) -- (7.5,0.75);
    \draw[thick] (8,0) .. controls (7.6,1) and (7.4,1) .. (7,0);
     \draw[thick] (5,0) .. controls (4.6,-1) and (4.4,-1) .. (4,0);
     \draw[->,thick] (4,0) --(4,1);
     \node at (4,1.25) {$S^z_j$};
     \node at (8.2,0) {$j$};
\end{tikzpicture}
    \caption{Cartoon depicting exponentially localized quarter spin in the ground state of $XXZ-\frac{1}{2}$ spin chain.}
    \label{fig:gsquar}
\end{figure}
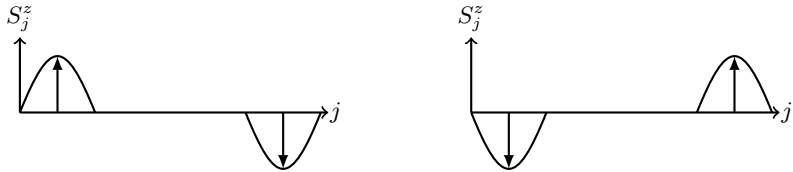

Such that the solution for the root density is immediate in the Fourier space
\begin{align}
    \tilde\rho_{\ket{0}}(\omega)&=\frac{(2N+1)e^{-\eta|\omega|}+2Ne^{-3\eta|\omega|}+(-1)^\omega e^{-\eta|\omega|}-2(-1)^\omega e^{-\eta|\omega|}-2(-1)^\omega e^{-3\eta|\omega|}-(1+(-1)^\omega)}{4\pi(1+e^{-4\eta|\omega|}+2e^{2\eta|\omega|})}\nonumber\\
    &- \frac{2(-1)^\omega e^{-5\eta|\omega|}+2(-1)^\omega e^{-3\eta|\omega|}}{{4\pi(1+e^{-4\eta|\omega|}+2e^{2\eta|\omega|})}}
\end{align}
The total number of center of bulk roots is given by
\begin{equation}
    M_{\ket{0}}=\int_{-\pi}^\pi \rho_{\ket{0}}(\lambda)\mathrm{d}\lambda=2\pi \tilde \rho_{\ket{0}}(0)=\frac{N-2}{2}
\end{equation}
And the total spin is given by
\begin{equation}
    S^z=N-2\left(1+\frac{N-2}{2}\right)=0
\end{equation}
The spin configuration is now made up of fractionalized spin$-\frac{1}{2}$ localized at the edges which point in opposite direction as shown the Fig.~\ref{fig:gsquar} 

We showed that $\pm\frac{1}{2}$ spin is localized at the edge of this fine tuned model. However, it is important to understand that any perturbations that do not close the gap and change the symmetry can change the boundary physics. Thus, we expect the regular $XXZ-1$ spin chain with Hamiltonian
\begin{equation}
    H_\Delta=\sum_{i=1}^{N=1}S^x_iS^x_{i+1}+S^y_iS^y_{i+1}+\Delta S^z_iS^z_{i+1}
\end{equation}
has to have this edge mode as long as $\Delta>\Delta_{c_2}$. We prove this claim numerically in the main text. Moreover, we also test by adding next near neighbor interaction or biquadratic term that as long as we are in the gapped antiferromagnetic regime, the edge modes exist as robust quantum observables.

\section{Spin-1 Bosonization}
Adapting the work of \cite{giamarchi2003quantum,schulz1986phase,timonen1985continuum}, we shall make the argument concrete for $S=1$ chain, the generalization to any higher spin being immediate as shown in \cite{giamarchi2003quantum,schulz1986phase}.

The low energy physics of a spin $1$ chain can be described by two {\it coupled} spin-$\frac{1}{2}$ chains in the following way 
\begin{equation}
    H_1=\sum_{\alpha=\{1,2\}}H_{\Delta,\frac{1}{2}}^\alpha+J_\perp \sum_j (\vec\sigma_{1,j}\cdot\vec\sigma_{2,j})_{\bar\Delta}
    \label{hamtobos}
\end{equation}
where $H_{\Delta,\frac{1}{2}}^\alpha$ is the single chain anisotropic Heisenberg  Hamiltonian, expressed in terms of spin-1/2 operators $\sigma_{\alpha,j}$ and $J_\perp$ is the interchain coupling. As discussed in detail in
\cite{timonen1985continuum,schulz1986phase,giamarchi2003quantum}, the Hamiltonian Eq.\eqref{hamtobos} can be bosonized in terms of the symmetric and and anti-symmetric field $\phi_\pm=\frac{\phi_1\pm\phi_2}{\sqrt{2}}$ where $\phi_1$ and $\phi_2$ corresponds to individual spin$-\frac{1}{2}$ chains and their respective dual variables $\theta_\pm$ as $H_1=H_++H_-$, where
\begin{align}
H_\pm = \int \frac{dx}{2\pi} \left[ u_\pm K_\pm (\partial_x\theta_\pm)^2 + \frac{u_\pm}{K_\pm} (\partial_x \phi_\pm)^2 \right]+  \int \frac{dx}{2\pi^2 
\alpha^2} \left[g_\pm\cos(\sqrt{8}\phi_\pm)+g_0 \xi_\mp\cos(\sqrt{2}\theta_-)\right],
\label{symham}
\end{align}
where $\xi_-=1$, $\xi_+=0$ and $\alpha$ is an arbitrary cutoff. 
Here $g_0=\pi J_\perp a$ and $g_\pm= J_\perp{\bar\Delta} a $, where $a$ is the lattice spacing. Likewise, the Luttinger liquid parameters $u_\pm,K_\pm$ for the symmetric and and anti-symmetric fields are related to those of the single chains $K = \frac{\pi}{2\arccos(-\Delta)}$, $u = \frac{K}{2K-1}\sin(\frac{\pi}{2K})$ as $u_\pm = u \left( 1 \pm \frac{KJ_\perp{\bar\Delta} a}{2 \pi u} \right)$, $K_\pm= K \left( 1 \mp \frac{KJ_\perp{\bar\Delta} a}{2 \pi u} \right)$.

The low energy  description of the spin-1 $XXZ$ chain is now given by $H_++H_-$ \cite{giamarchi2003quantum}. In the antiferromagnetic regime that  concerns, $g_+$ is relevant in the symmetric part of the Hamiltonian Eq.\eqref{symham} and hence it opens a mass gap in $H_+$ and the field $\phi_+$ develops long range antiferromagnetic order. Likewise, in the anti-symmetric part, the field $\phi_-$ also develops long range order while  the correlation in the dual field $\theta_-$ are exponentially decaying. Since the term $\cos(\sqrt{2}\theta_-)$ is irrelevant whereas $\cos(\sqrt{8}\phi_-)$ is relevant, we find that the low energy physics of the Hamiltonian $H_1$  is given by two copies of  $XXZ-\frac{1}{2}$ chain with an inter-chain coupling $g_0$ that is irrelevant. This substantiates our claim that the edge modes in the $XXZ-1$ chain are  of magnitude $\frac{1}{2}$, formed of the robust edge modes of magnitude $\frac{1}{4}$ in the two copies of the $XXZ-\frac{1}{2}$ chains. The three conditions mentioned in the introduction underlie our solution as the model is gapped and has both long range antiferromagnetic order and $U(1)$ symmetry  all of which are crucial to complete the RG argument discussed above.

\end{document}